\documentclass[reprint,aps,pra,showkeys,showpacs]{revtex4-1}  

\usepackage{graphicx}
\usepackage{epstopdf}
\usepackage{epsfig}
\usepackage{xcolor}   
\usepackage{amsmath}

\begin{document}

\title{Front-based construction of quantum droplets, bubbles, and hole states in a Bose mixture}
\author{Sherzod R. Otajonov$^{1, 2}$}
\author{Fatkhulla Kh. Abdullaev$^{1}$}
\affiliation{$^1$ Uzbekistan Academy of Sciences S. A. Azimov Physical-Technical Institute, Chingiz Aytmatov Str. 2-B, 100084, Tashkent, Uzbekistan}
\affiliation{$^2$ National University of Uzbekistan, Department of Theoretical Physics, 100174, Tashkent, Uzbekistan}

\begin{abstract}
We study nonlinear holes, bubbles, kink fronts, and quantum-droplet-like states in an elongated Bose mixture governed by an extended Gross-Pitaevskii equation with attractive cubic mean-field interactions and a repulsive quartic Lee-Huang-Yang contribution. We analytically classify the constant-amplitude backgrounds and show from the Bogoliubov-de Gennes spectrum that the upper branch is modulationally stable. In contrast, the lower branch is unstable to long-wavelength perturbations. The analysis of the grand-potential density and thermodynamic pressure allows us to determine the vacuum-finite-density coexistence point and the unique background supporting stationary kink and antikink fronts. Symmetric holes on the stable background are characterized by their density deficit and integral width, both of which diverge as the coexistence point is approached and the hole separates into two nonlinear fronts. A multiplicative kink-antikink construction connects localized bell-shaped states to broad flat-top droplets and approaches the stationary quantum-droplet family as the front separation increases. An antikink-kink sum, in which the two front profiles enter the superposition with the same sign, generates bubble-like density depletions without a phase jump.

In contrast, an antikink-kink difference, in which the two front profiles enter the superposition with opposite signs, produces dark-hole-like profiles with a $\pi$ phase difference between their asymptotic backgrounds. The bubble depth increases with front separation, providing a quantitative distinction between gray-like and dark-like profiles. Real-time simulations show that deep bubbles can persist under weak random perturbations over the simulated interval, demonstrating finite-time robustness without establishing spectral stability. In the asymmetric two-component model, localized relative-density modes can induce spinor-driven breakup even on a modulationally stable background, while dynamically robust holes persist near the kink limit. It is shown that collisions of kink-antikink droplet-like states display phase-dependent merging, asymmetric particle transfer, and effective reflection. These results connect isolated nonlinear fronts with bubbles, phase-jump holes, and self-bound droplet states in beyond-mean-field Bose mixtures.
\end{abstract}
\maketitle

\section{Introduction}
\label{intro}

The seminal works on ultradilute Bose mixtures established the possibility of stabilizing Bose-Einstein condensates (BECs) via quantum fluctuations \cite{Petrov2015, AP2016}. Beyond-mean-field effects associated with quantum fluctuations are commonly described by the Lee-Huang-Yang (LHY) correction to the energy of the condensate \cite{LHY}. When the residual mean-field interaction is sufficiently weak and comparable to the LHY contribution, quantum fluctuations may play a decisive role in determining the system's properties and stability. In three-dimensional two-component BECs, the LHY contribution provides an effective repulsive nonlinear term associated with an energy density proportional to $n^{5/2}$, where $n$ is the condensate density. This repulsion can compensate the residual mean-field attraction and prevent collapse.

One of the most important consequences of this competition is the formation of quantum droplets (QDs), which are self-bound states stabilized by the balance between residual mean-field attraction and the repulsion induced by quantum fluctuations \cite{Petrov2015}. Quantum droplets in Bose-Bose mixtures have been observed experimentally in Refs.~\cite{exp_bb1,exp_bb2,exp_bb3}. A related stabilization mechanism occurs in dipolar BECs, where the attractive part of the long-range dipole-dipole interaction is balanced by the repulsive LHY contribution. Quantum droplets in dipolar gases have likewise been observed experimentally \cite{exp_dip1,exp_dip2,exp_dip3,exp_dip4}. These developments have stimulated extensive investigations of the existence, stability, and dynamics of self-bound states governed by competing mean-field and beyond-mean-field nonlinearities.

An important feature of the LHY correction is its strong dependence on dimensionality and transverse confinement. In the strict one-dimensional limit, the effective LHY contribution changes both its functional form and sign relative to the three-dimensional case, and becomes attractive. In two dimensions, the corresponding beyond-mean-field correction depends logarithmically on density and may change character across different density regimes. The situation is different for elongated condensates that remain on the three-dimensional or dimensional-crossover side of the confinement problem. In this quasi-one-dimensional regime, the effective Gross-Pitaevskii equation may retain the repulsive quartic LHY nonlinearity characteristic of the three-dimensional equation of state \cite{Zin}. Thus, the physical properties of nonlinear excitations can differ substantially between the strict 1D limit and an elongated quasi-1D condensate~\cite{Otajonov2026_3}.

Although quantum droplets have been investigated extensively, other nonlinear states supported by the same competition of interactions have received considerably less attention. Kink-type states, which connect different asymptotic densities, and hole states, which represent localized density depletions embedded in a finite-density background, are of particular interest. The existence and stability of kinks and holes in low-dimensional Bose-Bose mixtures with LHY corrections have recently been studied in Refs.~\cite{Kartashov2022, Shukla2021, Katsimiga2023}. These studies revealed a close relation among self-bound droplets, kink fronts, and density-depleted nonlinear states. They also showed that, in a two-component description, the internal spinor degree of freedom may destabilize localized holes even when their homogeneous background remains modulationally stable.

The realization of the strict low-dimensional limits generally requires sufficiently strong transverse confinement \cite{Zin}. For experimentally relevant elongated condensates with weaker confinement, however, the effective beyond-mean-field equation may remain closer to its three-dimensional form. Such quasi-one-dimensional configurations have therefore attracted increasing attention. Beyond-mean-field Gross-Pitaevskii models for elongated Bose-Bose mixtures have been considered in Refs.~\cite{pak1,pak2, Otajonov2025, Otaj1, Otaj2,Marzug}, while related elongated dipolar systems were studied in Refs.~\cite{Edmonds, Bloch}. Beyond-mean-field effects in thick pancake geometries have also been considered in Refs.~\cite{Shamriz, Kit1, Kamchatnov}. These settings provide a natural framework for studying nonlinear structures generated by the competition between residual mean-field attraction and repulsive quantum-fluctuation effects.

In the present work, we investigate kink, antikink, hole, and quantum-droplet-like states in an elongated Bose-Bose mixture governed by a quasi-one-dimensional Gross-Pitaevskii equation containing an attractive cubic mean-field term and a repulsive quartic LHY contribution. We first determine the constant-amplitude backgrounds and analyze their modulational stability, identifying the branch that can support stable localized density depletions. We then formulate the stationary problem in terms of the energy, grand potential, and thermodynamic pressure and determine the vacuum-finite-density coexistence condition that selects the background supporting an isolated stationary kink. The properties of symmetric hole states and their limiting transformation into separated kink and antikink fronts are subsequently analyzed.

A central objective of the present study is to clarify the connection between isolated nonlinear fronts and self-bound droplet states. We show that a kink-antikink construction continuously connects compact bell-shaped localized states to broad flat-top droplet-like configurations, whereas reversing the ordering of the fronts produces gray- and dark-hole-like density depletions on a finite background. We further extend the analysis to an asymmetric two-component system, where the stability of localized holes depends on both total-density and relative-density excitations. Finally, we investigate collisions of kink-antikink droplet-like states and demonstrate their characteristic phase-dependent behavior, including merging, asymmetric particle transfer, and effective reflection.

The paper is organized as follows. In Sec.~\ref{sec:model}, we introduce the quasi-one-dimensional model. Sections~\ref{sec:constAmpl} and \ref{sec:MI} analyze the constant-amplitude backgrounds and their modulational stability, respectively. In Sec.~\ref{sec:KAK}, we derive the kink and antikink solutions and establish the vacuum-finite-density coexistence condition. Section~\ref{sec:Holes} examines symmetric hole states, kink-antikink droplet-like configurations, phase-jump dark-hole-like states, and same-phase bubble-like states. Section~\ref{sec:AsymHoles} extends the analysis to the asymmetric two-component system, addresses the stability of asymmetric holes, and investigates the collision dynamics of kink-antikink droplet-like states. Finally, Sec.~\ref{sec:Conc} summarizes the main results.

\section{Model Equation}
\label{sec:model}
We consider a symmetric two-component BEC with equal atomic masses and repulsive intracomponent interactions. The intercomponent interaction is attractive and is assumed to be tuned close to the repulsive one, so that the residual attractive mean-field contribution is small and the effective scalar description is applicable. The LHY correction includes beyond-mean-field contributions. In the symmetric approximation, where the same macroscopic wave function represents both condensate components, the coupled Gross-Pitaevskii system reduces to a single extended GPE.

In dimensionless form, the elongated one-dimensional model considered in this work is
\begin{equation}
i\frac{\partial \psi}{\partial t}
=
-\frac{1}{2}\frac{\partial^2\psi}{\partial x^2}
-\delta g |\psi|^2\psi
+
g_{\mathrm{LHY}} |\psi|^3\psi ,
\label{eq:gpe}
\end{equation}
where $\delta g>0$ and $g_{\mathrm{LHY}}>0$. The cubic term describes the attractive residual mean-field interaction, whereas the quartic LHY term is repulsive. 

Equation~(\ref{eq:gpe}) conserves the number of atoms (norm),
$$
 N=\int_{-\infty}^{+\infty}|\psi|^2\,dx
$$
the energy
\begin{equation*}
 E[\psi]=\int_{-\infty}^{+\infty}
 \left[
 \frac{1}{2}|\psi_x|^2
 -\frac{\delta g}{2}|\psi|^4
 +\frac{2g_{\mathrm{LHY}}}{5}|\psi|^5
 \right]dx.
\end{equation*}
and the momentum $M=(i/2)\int(\psi\psi_x^*-\psi^*\psi_x)\,dx$.

\section{Constant-Amplitude Backgrounds}
\label{sec:constAmpl}
The analysis of homogeneous backgrounds is required for two reasons. First, constant-amplitude states provide the reference configurations about which modulational instability is defined. Second, dark-hole, kink-like, and related localized structures are embedded in, or asymptotically approach, finite-density backgrounds. Therefore, before constructing such nonlinear excitations, it is necessary to determine which homogeneous branches exist and which of them are dynamically admissible. For a given chemical potential within the interval where two nonzero homogeneous solutions coexist, the algebraic background relation yields a lower-amplitude and an upper-amplitude branch. The lower branch is modulationally unstable, whereas the upper branch is modulationally stable and can therefore serve as the asymptotic background for dark-hole-type solutions. In particular, a modulationally unstable background cannot support a robust stationary hole under real-time evolution.

We first consider homogeneous stationary solutions of Eq.~(\ref{eq:gpe}) in the form
\begin{equation}
\psi(x,t)=a e^{-i\mu t},
\qquad
a\geq 0,
\label{eq:plane_wave}
\end{equation}
where $a$ is the background amplitude and $\mu$ is the chemical potential. The equation for $\mu$ is:
\begin{equation}
\mu
=
-\delta g\,a^2
+
g_{\mathrm{LHY}}a^3 .
\label{eq:mu_of_a}
\end{equation}
Equivalently, the background amplitude is determined by the cubic algebraic equation
\begin{equation}
g_{\mathrm{LHY}}a^3-\delta g\,a^2-\mu=0 .
\label{eq:amplitude_cubic}
\end{equation}
It is important to emphasize that Eq.~(\ref{eq:amplitude_cubic}) is cubic in the amplitude $a$. Hence, depending on $\mu$, it may have three real roots. However, not all algebraic roots correspond to distinct physical background amplitudes, because $a$ represents the modulus of the condensate wave function and must satisfy $a\geq 0$.

The turning point of the function $\mu(a)$ is found from
\begin{equation}
\frac{d\mu}{da}
=
-2\delta g\,a
+
3g_{\mathrm{LHY}}a^2 .
\end{equation}
Apart from the trivial point $a=0$, the nonzero critical amplitude is
\begin{equation}
a_{\mathrm{cr}}
=
\frac{2\delta g}{3g_{\mathrm{LHY}}}.
\label{eq:a_cr}
\end{equation}
The corresponding critical chemical potential is
\begin{equation}
\mu_{\mathrm{cr}}
=
-\frac{4(\delta g)^3}{27g_{\mathrm{LHY}}^2}.
\label{eq:mu_cr}
\end{equation}
For $\mu<\mu_{\mathrm{cr}}$, the cubic equation~(\ref{eq:amplitude_cubic}) has no positive real root, and therefore no nonzero physical homogeneous background exists. At $\mu=\mu_{\mathrm{cr}}$, two positive roots merge at $a=a_{\mathrm{cr}}$.

For
\begin{equation}
\mu_{\mathrm{cr}}\leq \mu\leq 0,
\end{equation}
the cubic equation has three real roots. To express them, we introduce
\begin{equation}
\theta(\mu)
=
\frac{1}{3}
\arccos
\left[
1+
\frac{27g_{\mathrm{LHY}}^2\mu}{2(\delta g)^3}
\right].
\label{eq:theta}
\end{equation}
The two nonnegative roots are
\begin{equation}
a_+(\mu)
=
\frac{\delta g}{3g_{\mathrm{LHY}}}
\left[
1+2\cos\theta(\mu)
\right],
\label{eq:a_plus}
\end{equation}
and
\begin{equation}
a_-(\mu)
=
\frac{\delta g}{3g_{\mathrm{LHY}}}
\left[
1+2\cos\left(\theta(\mu)-\frac{2\pi}{3}\right)
\right].
\label{eq:a_minus}
\end{equation}
Here $a_+(\mu)$ denotes the upper background branch, while $a_-(\mu)$ denotes the lower branch. These two branches coincide at $\mu=\mu_{\mathrm{cr}}$:
\begin{equation}
a_+(\mu_{\mathrm{cr}})
=
a_-(\mu_{\mathrm{cr}})
=
a_{\mathrm{cr}}.
\end{equation}
At $\mu=0$, they become
\begin{equation}
a_+(0)=\frac{\delta g}{g_{\mathrm{LHY}}},
\qquad
a_-(0)=0.
\end{equation}

The third real root in the interval $\mu_{\mathrm{cr}}\leq \mu\leq 0$ is
\begin{equation}
a_3(\mu)
=
\frac{\delta g}{3g_{\mathrm{LHY}}}
\left[
1+2\cos\left(\theta(\mu)-\frac{4\pi}{3}\right)
\right].
\label{eq:a_third}
\end{equation}
This root satisfies
\begin{equation}
a_3(\mu)\leq 0
\end{equation}
throughout the interval $\mu_{\mathrm{cr}}\leq \mu\leq 0$, with equality only at $\mu=0$. Therefore, it does not represent an additional positive background amplitude. If one works with a real signed stationary field, a negative asymptotic value may be interpreted as a phase-shifted representation of the same density. However, for the present constant-amplitude classification, $a$ is the modulus of the condensate wave function, and the third root is not counted as a separate physical branch.

For $\mu>0$, the cubic equation (\ref{eq:amplitude_cubic}) has only one positive real root. This root is the continuation of the upper branch and may be written in hyperbolic form as
\begin{equation}
a_+(\mu)
=
\frac{\delta g}{3g_{\mathrm{LHY}}}
\left[
1+2\cosh\eta(\mu)
\right],
\qquad
\mu>0,
\label{eq:a_plus_positive_mu}
\end{equation}
where
\begin{equation}
\eta(\mu)
=
\frac{1}{3}
\operatorname{arcosh}
\left[
1+
\frac{27g_{\mathrm{LHY}}^2\mu}{2(\delta g)^3}
\right].
\label{eq:eta}
\end{equation}
The other two roots are complex conjugates and have no direct interpretation as real background amplitudes. Thus, the upper branch extends from $\mu=\mu_{\mathrm{cr}}$ to positive chemical potentials, whereas the lower branch exists only for $\mu_{\mathrm{cr}}\leq \mu\leq 0$.

\section{Analysis of modulational instability}
\label{sec:MI}
We now analyze the modulational stability of the homogeneous backgrounds. This step is central to the present work because it determines which constant-amplitude branch can serve as a physically relevant background for localized hole-like states. It also identifies the parameter regimes in which small perturbations grow exponentially and may trigger the formation of localized density modulations. Therefore, the modulational-instability analysis connects the classification of homogeneous states with the subsequent study of nonlinear localized excitations and their dynamics.

Let a small complex perturbation $\xi(x,t)$ be added to the constant-amplitude state:
\begin{equation}
\psi(x,t)
=
\left[
a+\xi(x,t)
\right]e^{-i\mu t},
\qquad
|\xi|\ll a .
\label{eq:perturbed_background}
\end{equation}
Substituting Eq.~(\ref{eq:perturbed_background}) into Eq.~(\ref{eq:gpe}), using the background relation (\ref{eq:mu_of_a}), and retaining only terms linear in $\xi$ and $\xi^*$, we obtain
\begin{equation}
i\frac{\partial \xi}{\partial t}
+
\frac{1}{2}\frac{\partial^2 \xi}{\partial x^2}
+
\left(
\delta g\,a^2
-
\frac{3}{2}g_{\mathrm{LHY}}a^3
\right)
\left(
\xi+\xi^*
\right)
=
0 .
\label{eq:linearized_xi}
\end{equation}
This is the linearized Gross-Pitaevskii equation for small perturbations around the homogeneous state. It is equivalent to the Bogoliubov-de Gennes linearization for a spatially uniform background.

For compactness, we introduce
\begin{equation}
Q(a)
=
-\delta g\,a^2
+
\frac{3}{2}g_{\mathrm{LHY}}a^3 .
\label{eq:Q}
\end{equation}
Then Eq.~(\ref{eq:linearized_xi}) becomes
\begin{equation}
i\xi_t
=
-\frac{1}{2}\xi_{xx}
+
Q(a)
\left(
\xi+\xi^*
\right).
\label{eq:linearized_Q}
\end{equation}
The quantity $Q(a)$ determines the effective sign of the nonlinear response felt by long-wavelength perturbations. In particular, $Q(a)>0$ corresponds to modulational stability, whereas $Q(a)<0$ allows an unstable band of perturbation wavenumbers.

We seek normal-mode perturbations in the form
\begin{equation}
\xi(x,t)
=
u e^{ikx-i\Omega t}
+
v^* e^{-ikx+i\Omega^* t},
\label{eq:mi_ansatz}
\end{equation}
where $u$ and $v$ are infinitesimal perturbation amplitudes, $k$ is the modulation wavenumber, and $\Omega$ is the perturbation frequency. Substitution of Eq.~(\ref{eq:mi_ansatz}) into Eq.~(\ref{eq:linearized_Q}), 
the resulting dispersion relation is:
\begin{equation}
\Omega^2
=
\frac{k^2}{4}
\left[
k^2+4Q(a)
\right].
\label{eq:omega_squared}
\end{equation}
On the two physical branches $a=a_\pm(\mu)$, this becomes
\begin{equation}
\Omega_\pm(k,\mu)
=
\frac{k}{2}
\sqrt{
k^2
-
4\delta g\,a_\pm^2
+
6g_{\mathrm{LHY}}a_\pm^3
}.
\label{eq:omega_pm}
\end{equation}

The stability of each branch is determined by the sign of $Q(a)$. From Eq.~(\ref{eq:Q}),
\begin{equation}
Q(a)=a^2
\left(
-\delta g+\frac{3}{2}g_{\mathrm{LHY}}a
\right),
\end{equation}
so the sign changes at $a=a_{\mathrm{cr}}$. The upper branch satisfies $a_+>a_{\mathrm{cr}}$ except at the branch-merging point. Hence,
\begin{equation}
Q(a_+)>0,
\end{equation}
and Eq.~(\ref{eq:omega_squared}) remains real for all real $k$. The upper branch is therefore modulationally stable.

By contrast, the lower branch satisfies
\begin{equation}
0<a_-<a_{\mathrm{cr}},
\end{equation}
and hence
\begin{equation}
Q(a_-)<0.
\end{equation}
In this case, long-wavelength perturbations are unstable. The instability band is
\begin{equation}
0<k<k_{\mathrm{cr}},
\qquad
k_{\mathrm{cr}}
=
2\sqrt{-Q(a_-)} .
\label{eq:mi_band}
\end{equation}
Inside this interval, $\Omega_-$ is imaginary and the perturbation grows exponentially. The corresponding gain is
\begin{equation}
G(k,\mu)
\equiv
\operatorname{Im}\Omega_-
=
\frac{k}{2}
\sqrt{
-k^2-4Q(a_-)
}.
\label{eq:mi_growth}
\end{equation}
The gain vanishes at the edges of the instability band and attains its maximum at
\begin{equation}
k_{\max}
=
\frac{k_{\mathrm{cr}}}{\sqrt{2}}
=
\sqrt{-2Q(a_-)},
\qquad
G_{\max}
=
-Q(a_-).
\label{eq:kmax_gmax}
\end{equation}

These results show that the two physical amplitude branches have qualitatively different dynamical roles. The lower branch is modulationally unstable and therefore cannot provide a robust background for stationary hole-like excitations. The upper branch is modulationally stable and is the relevant finite-density background for the construction of holes, dark-notch states, and kink-related structures. This stability distinction is also important for numerical simulations: perturbations imposed on the lower branch are expected to grow and drive nonlinear pattern formation, whereas perturbations on the upper branch remain bounded in the linear regime.

\begin{figure}[htbp]
  \centerline{\includegraphics[width=4.45cm]{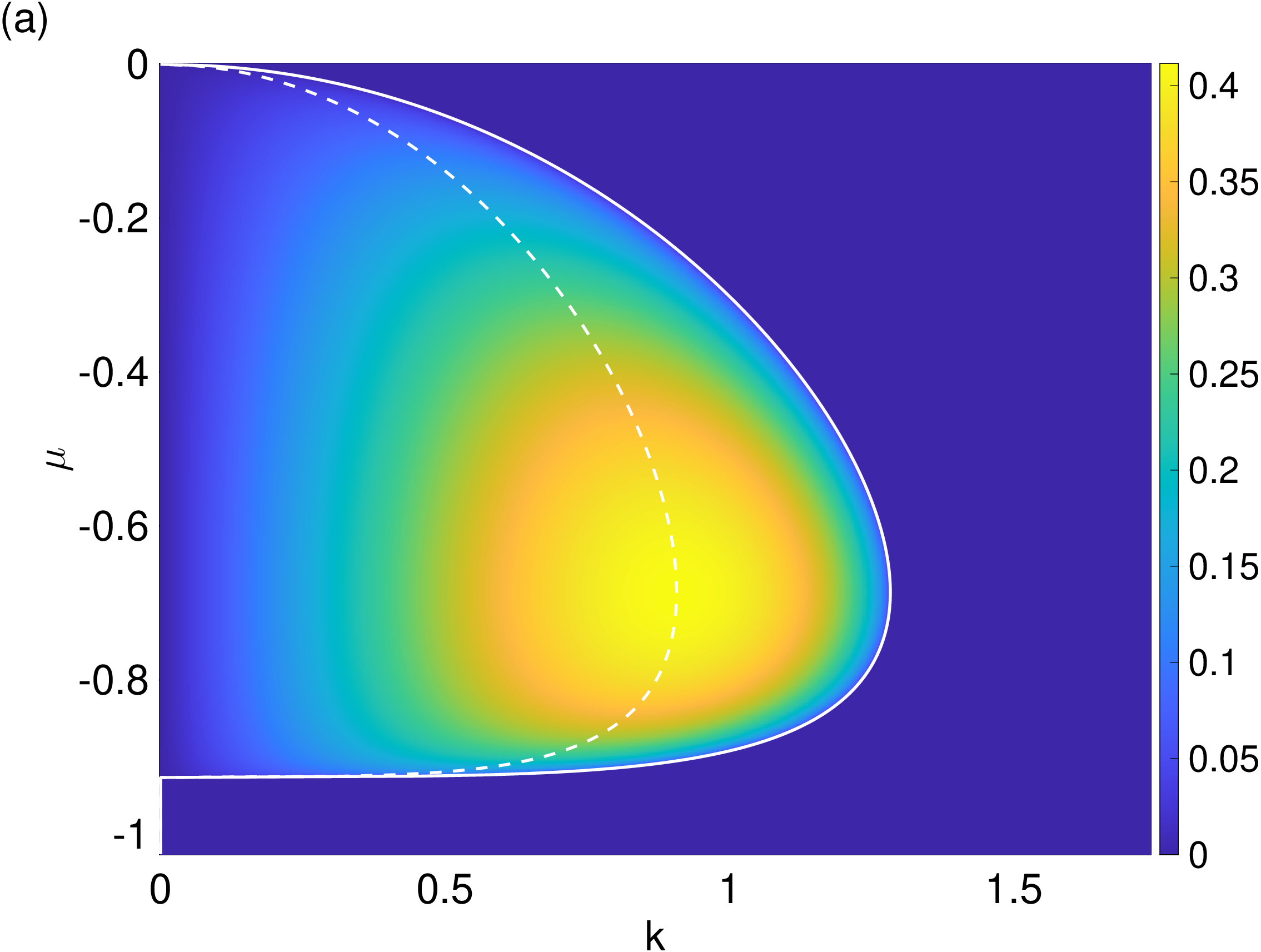}
  \includegraphics[width=4.45cm]{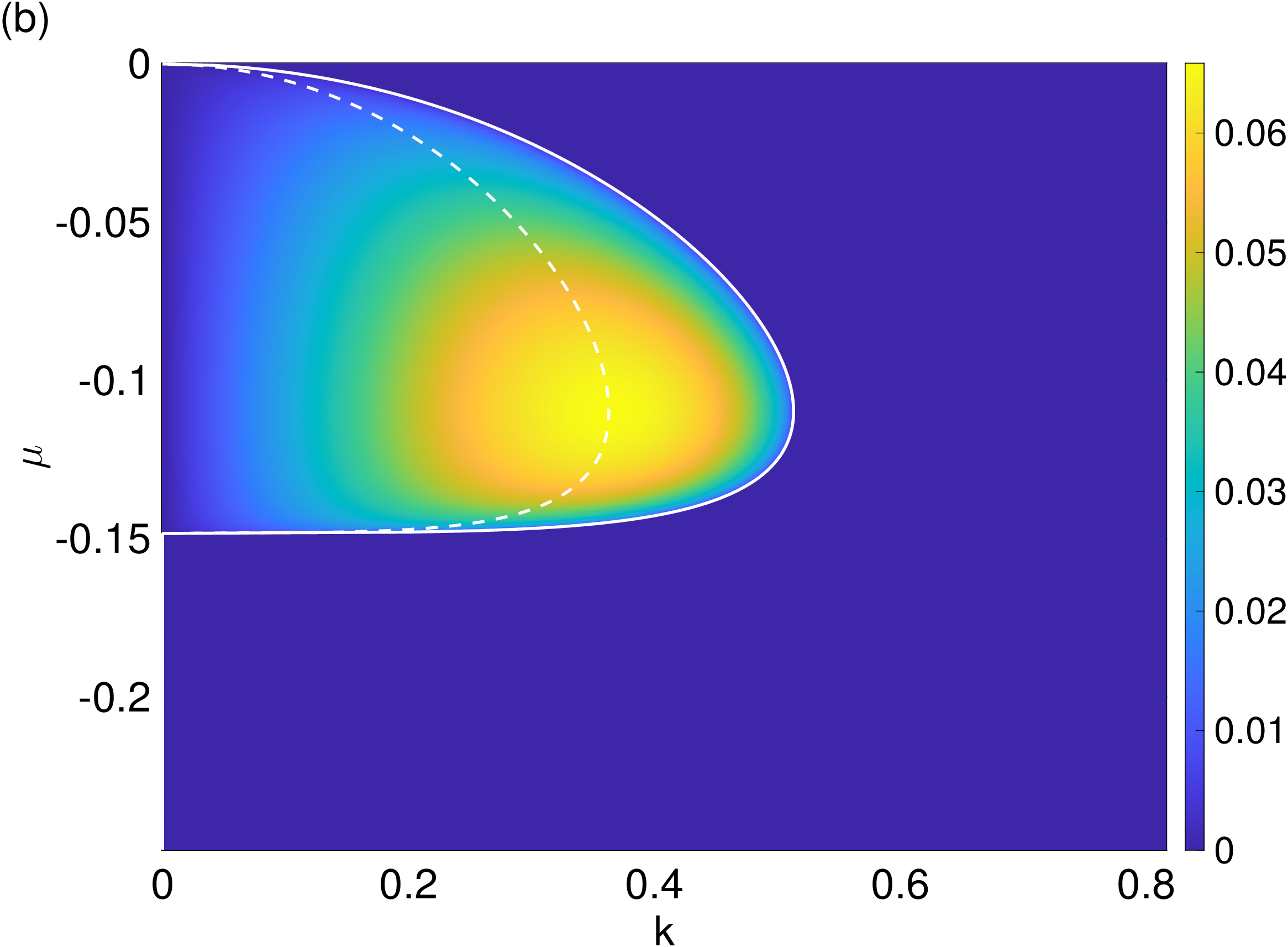}}
\caption{Modulational instability gain, $G=\operatorname{Im}(\Omega_-)$, as a function of the perturbation wavenumber $k$ and chemical potential $\mu$ for the lower constant-amplitude branch at fixed $\delta g=1$: (a) $g_{\mathrm{LHY}}=0.4$ and (b) $g_{\mathrm{LHY}}=1$.}
\label{fig-kk}
\end{figure}

Figures~\ref{fig-kk}(a) and \ref{fig-kk}(b) illustrate the modulational-instability gain $G(k,\mu)$ as a function of the chemical potential $\mu$ and perturbation wavenumber $k$ for two values of the LHY coefficient, $g_{\mathrm{LHY}}=0.4$ and $g_{\mathrm{LHY}}=1$, with $\delta g=1$ fixed. The gain spectra are shown for the lower constant-amplitude branch $a_-(\mu)$, since only this branch supports Bogoliubov modes with a nonzero imaginary part of the excitation frequency and can therefore exhibit modulational instability. In contrast, the upper constant-amplitude branch $a_+(\mu)$ satisfies $Q(a_+)>0$, for which the Bogoliubov excitation frequencies remain real for all real perturbation wavenumbers, and is therefore modulationally stable.

The unstable domain is bounded in both chemical potential and wavenumber. Along the chemical potential axis, the lower branch exists and is modulationally unstable only within the interval $\mu_{\mathrm{cr}}<\mu<0$, where $\mu_{\mathrm{cr}}$ denotes the branch-merging point. At the endpoints $\mu=\mu_{\mathrm{cr}}$ and $\mu=0$, the instability band collapses and the gain vanishes. For each fixed value of $\mu$ within this interval, the unstable wavenumbers satisfy,
$
0<k<k_{\mathrm{cr}},
k_{\mathrm{cr}}=2\sqrt{-Q(a_-)} ,
$
where $Q(a_- )<0$ on the lower branch. Because the gain is an even function of the wavenumber, the same instability occurs for negative wavenumbers within the symmetric interval
$
-k_{\mathrm{cr}}<k<0.
$
Therefore, only positive values of $k$ are displayed in Figs.~\ref{fig-kk}(a) and \ref{fig-kk}(b).

In the figures, the solid white curves represent the critical wavenumber $k_{\mathrm{cr}}(\mu)$, which bounds the modulationally unstable region, while the dashed white curves denote the most unstable wavenumber, at which the gain attains its maximum value. A comparison of Figs.~\ref{fig-kk}(a) and \ref{fig-kk}(b) reveals that increasing the quantum fluctuation coefficient $g_{\mathrm{LHY}}$ reduces both the extent of the instability domain and the peak gain values. This behaviour demonstrates the stabilising influence of the repulsive LHY correction, which counteracts the attractive cubic mean-field term responsible for the long-wavelength modulational instability.
$
k_{\max}(\mu)=\frac{k_{\mathrm{cr}}(\mu)}{\sqrt{2}},
$
at which the gain attains its maximum value. 
The modulational-instability analysis presented here serves a different purpose than in the previous study of the same quasi-one-dimensional cubic-quartic model. The conditions for MI, together with the linear growth stage and the subsequent nonlinear evolution, have already been analyzed in detail for both symmetric and asymmetric configurations in Ref.~\cite{Otajonov2025}. Therefore, the present work does not repeat the full MI analysis. Instead, we summarize only the results needed to discuss the nonlinear stationary states. In particular, when the homogeneous solutions are organized into lower- and upper-amplitude branches as functions of the chemical potential, the upper branch is modulationally stable, whereas the lower branch is modulationally unstable. This distinction is essential for the following analysis because the stable upper branch provides the physically relevant asymptotic background for hole states and kink-related structures.

\section{Stationary reduction and kink, antikink  solutions}
\label{sec:KAK}

Let us find kink, antikink and hole solutions by seeking stationary states of the governing equation in the form
$$
\psi(x,t)=q(x)e^{-i\mu t},
$$
where $\mu$ denotes the chemical potential. For the states considered below, the stationary profile $q(x)$ can be chosen to be real. Substituting the stationary ansatz into the governing equation yields
$$
\mu q
=
-\frac{1}{2}q_{xx}
-\delta g\,q^3
+
g_{\mathrm{LHY}}|q|^3q .
$$
The absolute value in the LHY term must be retained when the profile changes sign, as occurs for hole solutions that connect backgrounds with a relative phase shift of $\pi$.
The stationary equation may also be obtained by extremizing the grand potential functional at fixed chemical potential. The energy and norm functionals are
$$
E[q]
=
\int_{-\infty}^{+\infty}
\left[
\frac{1}{2}q_x^2
-\frac{\delta g}{2}q^4
+\frac{2g_{\mathrm{LHY}}}{5}|q|^5
\right]dx,
$$
$$
N[q]
=
\int_{-\infty}^{+\infty}q^2\,dx .
$$
The corresponding grand potential is
$$
\Omega[q,\mu]
=
E[q]-\mu N[q]
=
\int_{-\infty}^{+\infty}
\left[
\frac{1}{2}q_x^2
+\omega(q^2,\mu)
\right]dx ,
$$
where, in terms of the density $n=q^2$, the bulk energy density and the bulk grand potential density are
$$
\varepsilon(n)
=
-\frac{\delta g}{2}n^2
+\frac{2g_{\mathrm{LHY}}}{5}n^{5/2},
\quad
\omega(n,\mu)
=
\varepsilon(n)-\mu n .
$$
The transformation from $\varepsilon(n)$ to $\omega(n,\mu)$ is the Legendre transform of the energy density with respect to the conserved density $n$. It replaces the description at fixed particle number by the thermodynamically equivalent description at fixed chemical potential. The stationary equation is equivalently obtained from $\delta\Omega / \delta q=0$.

The thermodynamic pressure associated with a homogeneous state is
$$
P(n)
=
n\frac{\partial\varepsilon}{\partial n}
-\varepsilon(n)
=
-\frac{\delta g}{2}n^2
+\frac{3g_{\mathrm{LHY}}}{5}n^{5/2}.
$$
For a homogeneous stationary background, $\mu=\partial\varepsilon/\partial n$, and hence
$$
P(n)
=
\mu n-\varepsilon(n)
=
-\omega(n,\mu).
$$
Thus, equality of grand potential densities is equivalent to equality of pressures.

For the positive part of a kink profile, set $q(x)=g(x)\geq0$. The stationary equation then reduces to
$$
\mu g
=
-\frac{1}{2}g_{xx}
-\delta g\,g^3
+
g_{\mathrm{LHY}}g^4 .
$$
The first integral is:
$$
C
=
-\frac{1}{4}g_x^2
+\frac{1}{2}\omega(g^2,\mu).
$$
This relation can be interpreted as an energy-conservation law for the equivalent spatial dynamical system, where $x$ serves as the evolution variable. A stationary front connects two asymptotic states only when both correspond to the same value of this first integral. Equivalently, the vacuum and finite-density phases must have equal grand potential densities, so neither phase expands at the expense of the other, and a stationary interface can connect them.

At a homogeneous asymptotic state, where $g_x=0$, the first integral reduces to
$$
C
=
\frac{1}{2}\omega(g^2,\mu)
=
-\frac{1}{2}P(g^2).
$$
Consequently, equality of the first-integral constants at the two ends of a stationary front is precisely the condition of equal grand potential density, or equivalently equal thermodynamic pressure, in the two asymptotic phases.

A kink connects the vacuum to a nonzero constant-amplitude background. The corresponding boundary conditions are
$$
g(-\infty)=0,
\qquad
g(+\infty)=A_{\mathrm{co}},
\qquad
g_x(\pm\infty)=0 .
$$
At the vacuum boundary, the first integral yields $C=0$. It therefore becomes
$$
g_x^2
=
-2\mu g^2
-\delta g\,g^4
+\frac{4g_{\mathrm{LHY}}}{5}g^5 .
$$
For a real monotonic kink, the right-hand side must remain nonnegative throughout the interval $0\leq g\leq A_{\mathrm{co}}$ and must vanish at both asymptotic states.

Because both the energy density and particle density vanish in the vacuum, its grand potential density and pressure also vanish:
$$
\varepsilon(0)
=
\omega(0,\mu)
=
P(0)
=
0.
$$
Therefore, the condition $C=0$ implies that the finite-density state connected to the vacuum must satisfy
$$
\omega(A_{\mathrm{co}}^2,\mu_{\mathrm{co}})
=
-P(A_{\mathrm{co}}^2)
=
0.
$$
This is the thermodynamic coexistence condition between the vacuum and liquid-like finite-density phases.

At the finite-density boundary, where $g=A_{\mathrm{co}}$ and $g_x=0$, the first integral gives
$$
-2\mu A_{\mathrm{co}}^2
-\delta g\,A_{\mathrm{co}}^4
+\frac{4g_{\mathrm{LHY}}}{5}A_{\mathrm{co}}^5
=0 .
$$
The same asymptotic state must also satisfy the homogeneous-background relation
$$
\mu
=
-\delta g\,A_{\mathrm{co}}^2
+
g_{\mathrm{LHY}}A_{\mathrm{co}}^3 .
$$
Substituting the homogeneous-background relation into the first-integral condition yields
$$
A_{\mathrm{co}}^4
\left(
\delta g-\frac{6}{5}g_{\mathrm{LHY}}A_{\mathrm{co}}
\right)
=0 .
$$
In addition to the trivial vacuum solution, the nonzero coexistence amplitude is therefore
$$
A_{\mathrm{co}}
=
\frac{5\delta g}{6g_{\mathrm{LHY}}}.
$$
The corresponding chemical potential follows from the homogeneous-background relation:
\begin{equation}
\mu_{\mathrm{co}}
=
-\frac{25(\delta g)^3}{216g_{\mathrm{LHY}}^2}.
\label{eq:mu_co}
\end{equation}

The same coexistence amplitude follows directly from the zero-pressure condition,
$$
P(A_{\mathrm{co}}^2)
=
A_{\mathrm{co}}^4
\left(
-\frac{\delta g}{2}
+\frac{3g_{\mathrm{LHY}}}{5}A_{\mathrm{co}}
\right)
=0,
$$
which again yields $A_{\mathrm{co}}=5\delta g/(6g_{\mathrm{LHY}})$. At this amplitude, the Legendre-transformed energy density of the finite-density phase is equal to that of the vacuum:
$$
\omega(A_{\mathrm{co}}^2,\mu_{\mathrm{co}})
=
\varepsilon(A_{\mathrm{co}}^2)
-\mu_{\mathrm{co}}A_{\mathrm{co}}^2
=
0.
$$

The value $\mu_{\mathrm{co}}$ is the chemical potential at vacuum-finite-density coexistence. At this point, the vacuum and finite-density phases have equal grand-potential densities, equivalently zero pressure difference, so neither phase is thermodynamically favored over the other, and a stationary interface can connect them. Away from coexistence, the asymptotic states are not energetically degenerate, and an isolated stationary front connecting the vacuum to the finite-density phase cannot exist. Thus, the coexistence amplitude lies on the upper constant-amplitude branch, with $A_{\mathrm{co}}>a_{\mathrm{cr}}$ and $\mu_{\mathrm{cr}}<\mu_{\mathrm{co}}<0$, showing that the finite-density state supporting the kink is modulationally stable. The branch-merging point $\mu_{\mathrm{cr}}$ and the coexistence point $\mu_{\mathrm{co}}$ therefore have distinct physical meanings: the former marks the minimum chemical potential at which nonzero homogeneous states exist, whereas the latter selects the unique homogeneous background that can coexist with the vacuum through a stationary kink or antikink.

At the coexistence chemical potential, the stationary equation yields a heteroclinic front connecting the vacuum state to the finite-density background. This front constitutes a single kink. Introducing the shifted coordinate $X=x-x_0$, where $x_0$ denotes the kink center, allows the kink profile to be expressed in parametric form as follows:
$$
g_{\mathrm{K}}(X)
=
\frac{A_{\mathrm{co}}}{2}
\left[
s^2(X)-1
\right],
$$
where the auxiliary function $s(X)$ is determined implicitly by
$$
\ln\left(
\frac{s-1}{s+1}
\right)
+
\frac{1}{\sqrt{3}}
\ln\left(
\frac{\sqrt{3}+s}{\sqrt{3}-s}
\right)
=
\frac{5(\delta g)^{3/2}}
{6\sqrt{3}\,g_{\mathrm{LHY}}}
X .
$$
Here, $1<s<\sqrt{3}$. As $X\rightarrow-\infty$, $s\rightarrow1$ and $g_{\mathrm{K}}\rightarrow0$, while as $X\rightarrow+\infty$, $s\rightarrow\sqrt{3}$ and $g_{\mathrm{K}}\rightarrow A_{\mathrm{co}}$. Consequently, the solution connects the vacuum to the coexistence background monotonically.

The corresponding time-dependent kink solution is
$$
\Psi_{\mathrm{kink}}(x,t)
=
g_{\mathrm{K}}(x-x_0)e^{-i\mu_{\mathrm{co}}t}.
$$
The antikink is obtained by spatial reflection of the kink profile,
$$
\Psi_{\mathrm{antikink}}(x,t)
=
g_{\mathrm{K}}(x_0-x)e^{-i\mu_{\mathrm{co}}t}.
$$
It connects the finite-density coexistence background to the vacuum. Thus, the kink and antikink satisfy the asymptotic conditions $g_{\mathrm{K}}(-\infty)=0$, $g_{\mathrm{K}}(+\infty)=A_{\mathrm{co}}$, $g_{\mathrm{AK}}(-\infty)=A_{\mathrm{co}}$, and $g_{\mathrm{AK}}(+\infty)=0$.

The parameter $x_0$ determines only the position of the front and does not influence its intrinsic shape. The intrinsic width of the front is governed by the ratio of the attractive mean-field coefficient to the repulsive LHY coefficient. These kink and antikink fronts serve as the fundamental building blocks for the localized droplet-like and hole-like states discussed in the following sections.

\begin{figure}[htbp] 
  \centerline{\includegraphics[width=4.45cm]{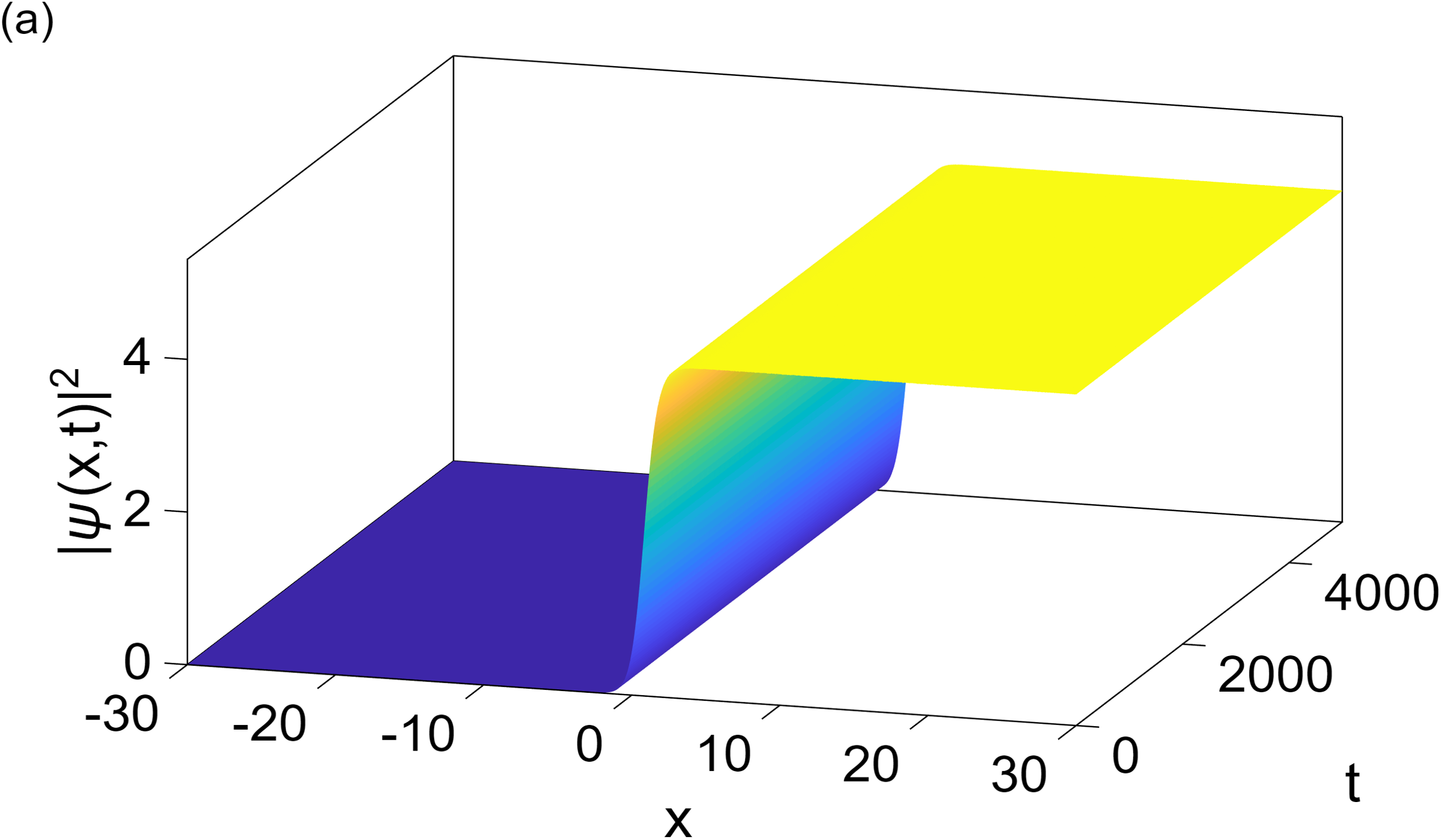} 
  \includegraphics[width=4.45cm]{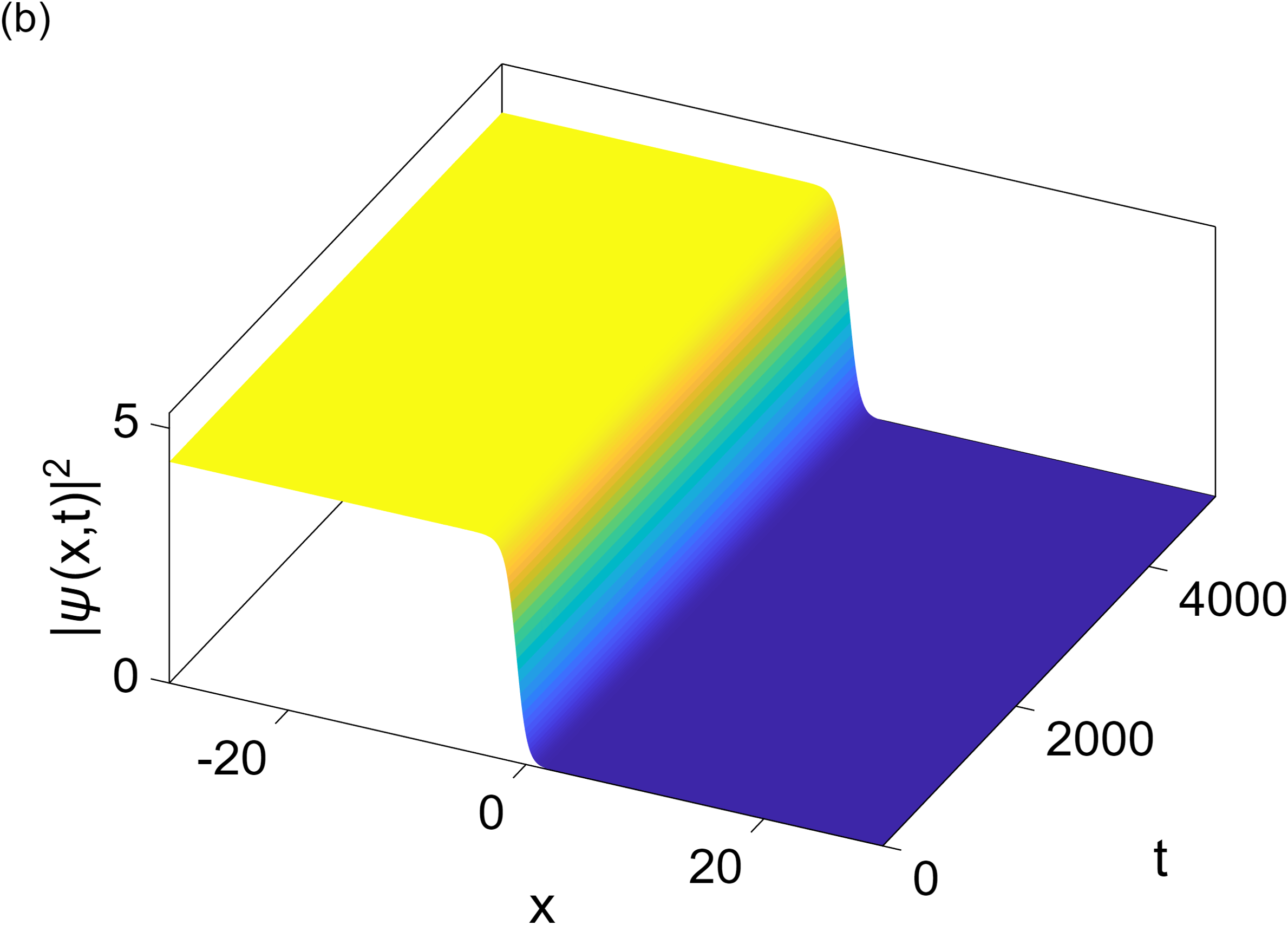}}
\caption{Stable evolution of the kink (a) and antikink (b) solutions at $\mu=-0.72$ under small random perturbations. The perturbation amplitude is $\epsilon=10^{-3}$, and the remaining parameters are $\delta g=1$ and $g_{\mathrm{LHY}}=0.4$. }
\label{fig-dynSymKAK}
\end{figure}

Figure~\ref{fig-dynSymKAK} shows the typical time evolution of the kink and antikink solutions in panels (a) and (b), respectively, after a small random perturbation is added to the initial stationary profiles. To assess dynamical stability, we define the initial conditions as follows:
$
\psi(x,0)=\psi_0(x)\left[1+\epsilon R(x)\right],
$
here, $\psi_0(x)$ denotes the stationary kink or antikink profile, $R(x)$ is a random function uniformly distributed over $[-1,1]$, and $\epsilon=10^{-3}$ represents the perturbation amplitude. The simulations indicate that both the kink and antikink maintain their shapes throughout the evolution, without growing oscillations or radiation-induced decay, confirming the robustness of these states against weak random perturbations.

\section{Hole and Bubble solutions}
\label{sec:Holes}

Now we will consider stationary hole states embedded in a finite-density background. These states are represented as
$$
\Psi_{\mathrm{h}}(x,t)=q(x)e^{-i\mu t},
$$
where $q(x)$ is real and sign changing. A symmetric hole centered at the origin satisfies
$$
q(-\infty)=-a,
\qquad
q(+\infty)=a,
\qquad
q(0)=0.
$$
The two asymptotic states have the same density but differ by a phase shift of $\pi$. The background amplitude is selected from the modulationally stable upper branch, $a=a_+(\mu)$, and satisfies the homogeneous relation $\mu=-\delta g\,a^2+g_{\mathrm{LHY}}a^3$. This stability condition fixes the background used below.

Because the two asymptotic states have the same density $n=a^2$, they possess identical energy density, grand potential density, and pressure:
$$
\varepsilon[q^2(-\infty)]
=
\varepsilon[q^2(+\infty)],
\,\,\,
\omega[q^2(-\infty),\mu]
=
\omega[q^2(+\infty),\mu],
$$
$$
P[q^2(-\infty)]
=
P[q^2(+\infty)]
=
P(a^2).
$$
Unlike the vacuum finite density kink, a stationary hole therefore does not require the background pressure itself to vanish. It only requires equal pressure on the two sides, which is automatically satisfied because the asymptotic densities are equal.

For a hole on an infinite nonzero background, both $E$ and $\Omega$ contain an infinite bulk contribution. The relevant finite thermodynamic quantity is consequently the excess grand potential relative to the homogeneous background,
$$
\Delta\Omega_{\mathrm{h}}
=
\int_{-\infty}^{+\infty}
\left[
\frac{1}{2}q_x^2
+
\omega(q^2,\mu)
-
\omega(a^2,\mu)
\right]dx .
$$
This quantity measures the grand potential cost of creating the localized density depletion and its associated phase change within the uniform background.

For a symmetric hole, it is sufficient to consider the half-line $x\geq0$, where $q(x)=g(x)\geq0$, $g(0)=0$, and $g(+\infty)=a$. On this half-line, the profile obeys the same positive stationary equation as the kink. Applying the first integral and imposing the finite-density boundary condition at infinity yields
$$
g_x^2
=
\delta g\left(a^4-g^4\right)
-\frac{4g_{\mathrm{LHY}}}{5}
\left(a^5-g^5\right)
+
2\mu\left(a^2-g^2\right).
$$
This expression determines the spatial profile of the hole for $x\geq0$. The complete solution is obtained by odd continuation, $q(-x)=-q(x)$.

The slope at the center follows by setting $g=0$:
$$
g_x^2(0)
=
a^4
\left(
\frac{6}{5}g_{\mathrm{LHY}}a-\delta g
\right)
=
2P(a^2).
$$
Thus, the central slope is directly controlled by the thermodynamic pressure of the background. A finite-width hole requires $P(a^2)>0$, whereas the vanishing-pressure condition $P(a^2)=0$ marks the transition to infinitely separated interfaces.

Therefore, a finite-width hole requires the modulationally stable upper-branch background to satisfy $a>A_{\mathrm{co}}$. When $a=A_{\mathrm{co}}$, or equivalently $\mu=\mu_{\mathrm{co}}$, the central slope vanishes and the hole width diverges. In this regime, the hole ceases to represent a finite localized notch and instead approaches a pair of separated nonlinear fronts between stable backgrounds, with an extended vacuum region between them.

At this limit, the excess bulk grand potential density of the vacuum region relative to the finite-density background vanishes:
$$
\omega(0,\mu_{\mathrm{co}})
-
\omega(A_{\mathrm{co}}^2,\mu_{\mathrm{co}})
=
0.
$$
Consequently, increasing the separation between the two fronts does not produce a bulk grand potential cost. This thermodynamic degeneracy explains why the width of the hole can diverge at coexistence.

This observation explains the relationship between holes and kink states. A finite-width hole corresponds to a sign changing density depletion on the modulationally stable upper background. As the coexistence limit is approached, the depletion broadens continuously and separates into two asymptotically distinct interfaces. Thus, kink and antikink solutions appear as limiting cases within the hole family, whereas finite holes represent bound front-pair configurations on the stable upper background.

\begin{figure}[htbp]
  \centerline{\includegraphics[width=4.45cm]{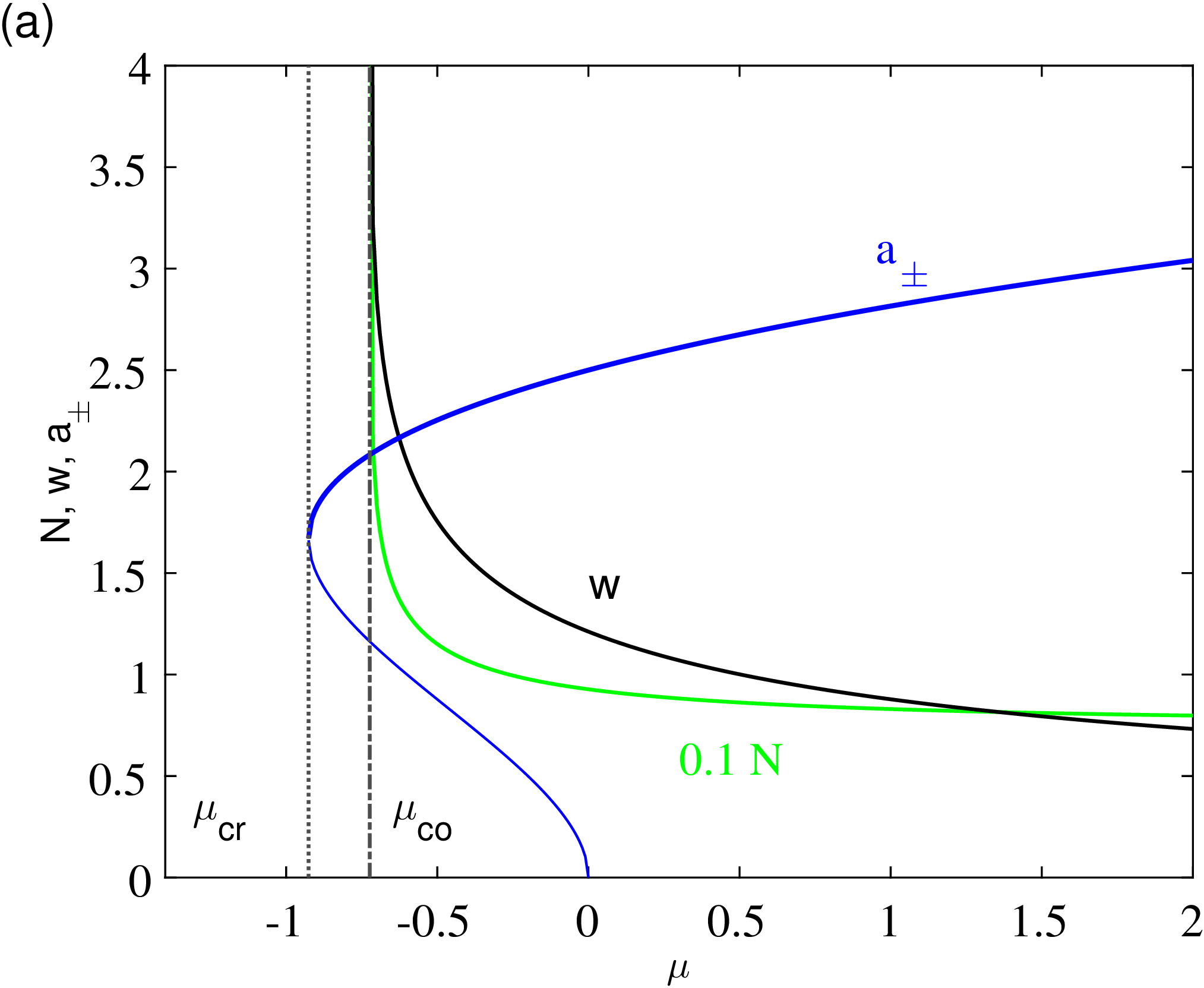}
  \includegraphics[width=4.45cm]{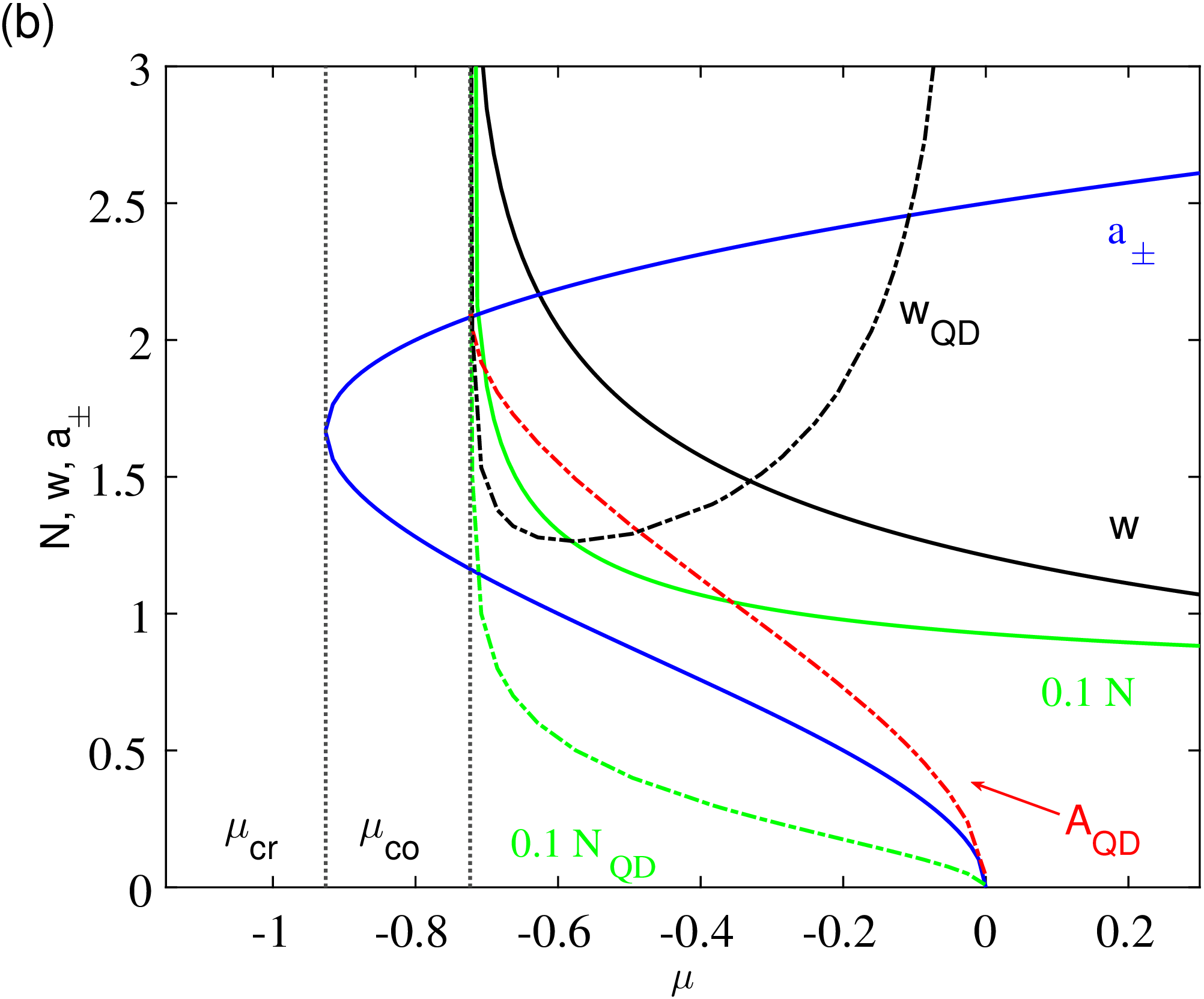}}
\caption{(a) Parameters of 1D symmetric holes versus the chemical potential $\mu$. The left and right vertical dotted lines mark $\mu_{\mathrm{cr}}$ [Eq.~(\ref{eq:mu_cr})] and $\mu_{\mathrm{co}}$ [Eq.~(\ref{eq:mu_co})], respectively. The green curve shows the redefined norm $N(\mu)$, scaled by $0.1$ for clarity, while the black curve gives the notch width $w(\mu)$. The blue curves represent the constant-background branches $a_{\pm}(\mu)$, with the thick upper and thin lower segments corresponding to $a_{+}(\mu)$ and $a_{-}(\mu)$, respectively. (b) Zoom of panel (a), including the norm $N_{\mathrm{QD}}(\mu)$, amplitude $A_{\mathrm{QD}}(\mu)$, and width $w_{\mathrm{QD}}(\mu)$ of symmetric self-bound 1D quantum droplets, obtained from a super-Gaussian function-based variational approach. Other parameters are $\delta g=1$ and $g_{\mathrm{LHY}}=0.4$.}
\label{fig-parSymHoles}
\end{figure}

Modulational stability of the upper constant-amplitude branch is required for robust hole states. Since hole solutions are embedded in a finite-density background, their asymptotic states must be modulationally stable. The upper branch $a_+(\mu)$ satisfies this criterion, whereas the lower branch $a_-(\mu)$ is modulationally unstable. Therefore, the following analysis constructs hole solutions on the stable upper background.

\begin{figure}[htbp]
  \centerline{\includegraphics[width=4.45cm]{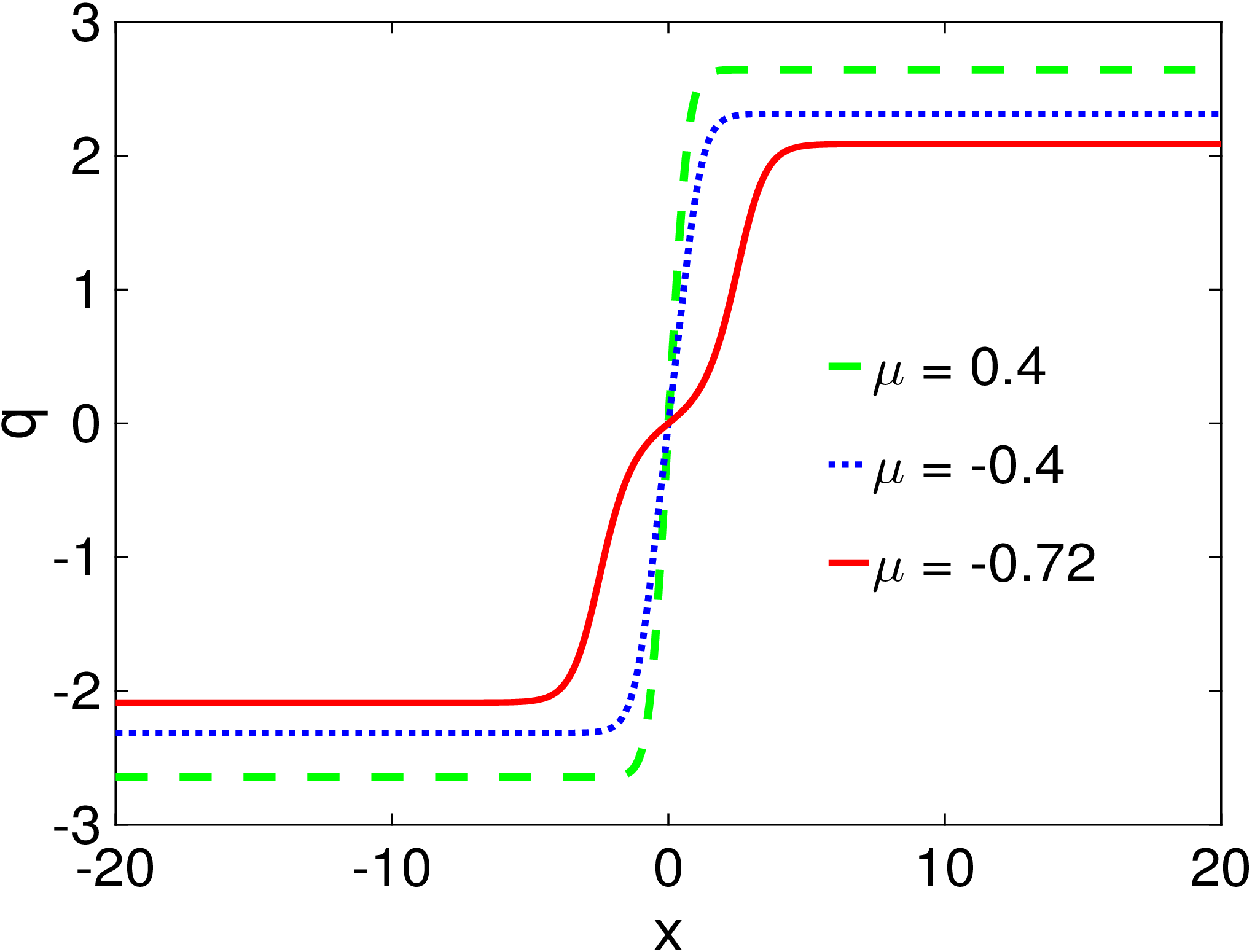}}
\caption{Representative hole profiles for different values of the chemical potential $\mu$. From top to bottom, the curves correspond to $\mu=0.4$ (green dashed line), $\mu=-0.4$ (blue dotted line), and $\mu=-0.72$ (red solid line). Other parameters are $\delta g=1$ and $g_{\mathrm{LHY}}=0.4$.}
\label{fig-profSymHoles}
\end{figure}

The main characteristics of the symmetric hole family are summarised in Figs.~\ref{fig-parSymHoles}(a) and \ref{fig-parSymHoles}(b), and representative density profiles are shown in Fig.~\ref{fig-profSymHoles}. As described in Ref.~\cite{Kartashov2022}, the hole solutions are described by the background amplitude $a_+(\mu)$ and the redefined norm $N(\mu)$, and the integral width of the density notch, $w(\mu)$,
\begin{equation}
N(\mu)
=
\int_{-\infty}^{+\infty}
\left[
a_+^2(\mu)-|\psi(x)|^2
\right] dx ,
\label{eq:hole_norm}
\end{equation}
and the integral width of the density notch,
\begin{equation}
w(\mu)
=
2
\left[
\frac{
\int_{-\infty}^{+\infty}
x^2\left[
a_+^2(\mu)-|\psi(x)|^2
\right] dx
}{
N(\mu)
}
\right]^{1/2}.
\label{eq:hole_width}
\end{equation}
$N(\mu)$ quantifies the density deficit relative to the homogeneous upper-branch background, whereas $w(\mu)$ measures the spatial scale of the notch.

Figure~\ref{fig-parSymHoles}(a) illustrates the dependence of $N(\mu)$, $w(\mu)$, and the constant-background amplitudes $a_{\pm}(\mu)$ on the chemical potential. The left vertical dotted line indicates the critical value $\mu_{\mathrm{cr}}$, where the two constant-amplitude branches merge and below which no nonzero homogeneous background exists. The right vertical dotted line denotes the cutoff value $\mu_{\mathrm{co}}$, where the hole family approaches the kink limit. Thus, $\mu_{\mathrm{cr}}$ marks the disappearance of constant-amplitude backgrounds, while $\mu_{\mathrm{co}}$ marks the transformation of holes into a pair of separated kink-like fronts.

A notable property of these holes is that they exist for both negative and positive chemical potentials. This differs from standard dark solitons of the repulsive cubic nonlinear Schr\"odinger equation, where the homogeneous background typically corresponds to $\mu>0$. In the cubic-quartic model considered here, the LHY correction modifies the finite-density background, allowing hole states to persist outside the interval where self-bound droplets are present.

The width $w(\mu)$ decreases as the chemical potential increases, indicating that the density notch becomes narrower on higher-density backgrounds. In contrast, the redefined norm $N(\mu)$ is nonmonotonic. Both $N(\mu)$ and $w(\mu)$ diverge as $\mu$ approaches $\mu_{\mathrm{co}}$ from above, showing the separation of the two density fronts that form the hole. In this regime, the hole ceases to be a localized notch and becomes a pair of out-of-phase kink and antikink structures.

Representative profiles in Fig.~\ref{fig-profSymHoles} demonstrate this transformation. For positive chemical potential, the hole remains relatively narrow and strongly localized on the stable background. As $\mu$ decreases toward $\mu_{\mathrm{co}}$, the notch broadens and the profile gradually evolves into two well-separated fronts. Therefore, the cutoff $\mu_{\mathrm{co}}$ indicates the transition between finite-width hole states and isolated kink-type solutions.

Figure~\ref{fig-parSymHoles}(b) further illustrates the relationship between the hole family and symmetric self-bound quantum droplets. The droplet norm $N_{\mathrm{QD}}(\mu)$, amplitude $A_{\mathrm{QD}}(\mu)$, and width $w_{\mathrm{QD}}(\mu)$ are derived from a super-Gaussian trial function-based variational approximation and are plotted alongside the hole characteristics. The self-bound droplet family lies within the same interval, $\mu_{\mathrm{co}}\leq \mu \leq 0$.
At $\mu=\mu_{\mathrm{co}}$, the droplet approaches the flat-top limit, where its norm and width diverge, and its plateau amplitude coincides with the background amplitude of the hole solution. Thus, the same limiting chemical potential links two complementary nonlinear states. In this limit, the hole becomes an extended kink-antikink configuration on a finite background, while the self-bound droplet becomes a flat-top state that can be interpreted as a localized structure formed by a kink-antikink pair.

\begin{figure}[htbp]
  \centerline{\includegraphics[width=4.45cm]{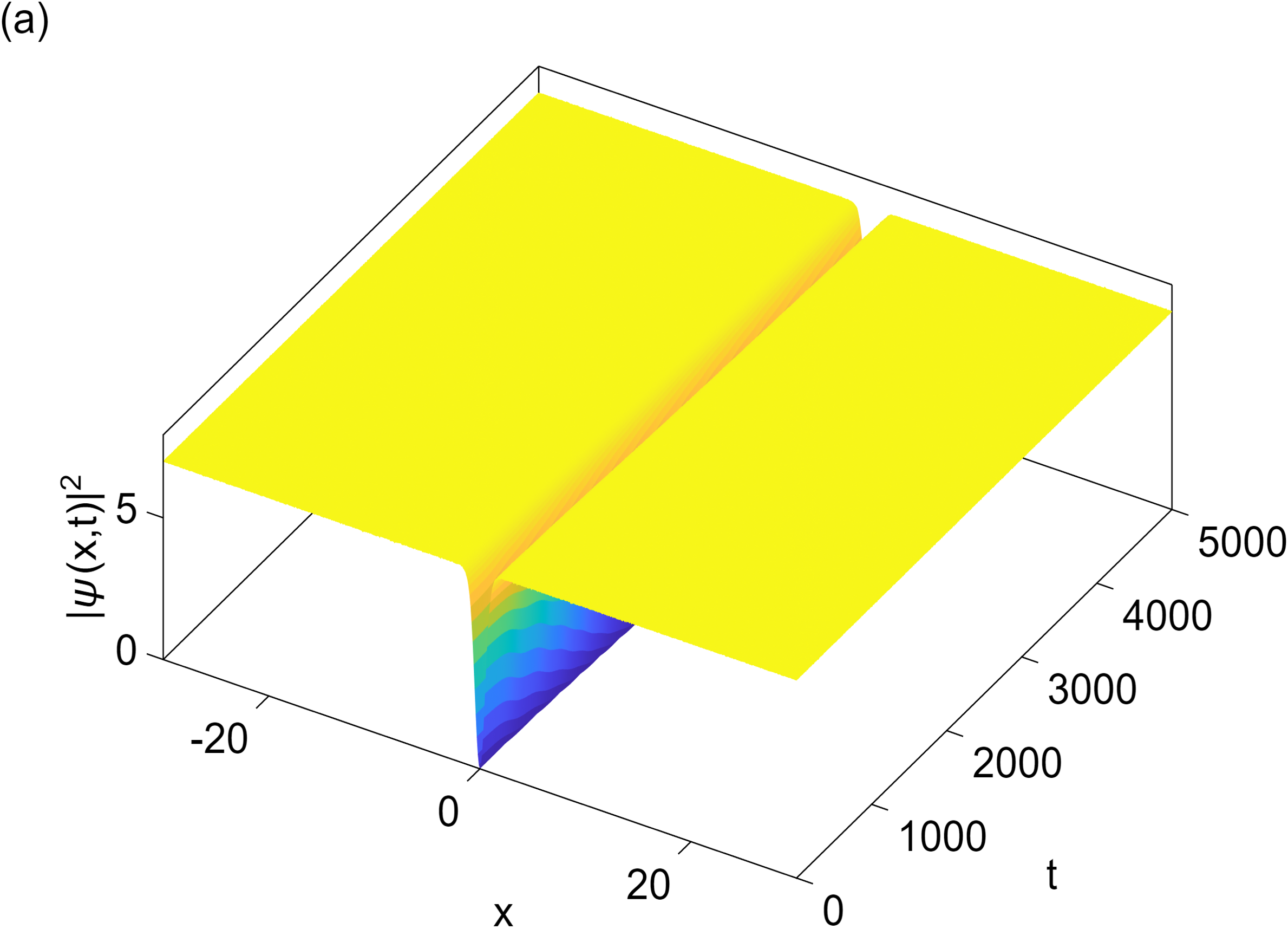}
  \includegraphics[width=4.45cm]{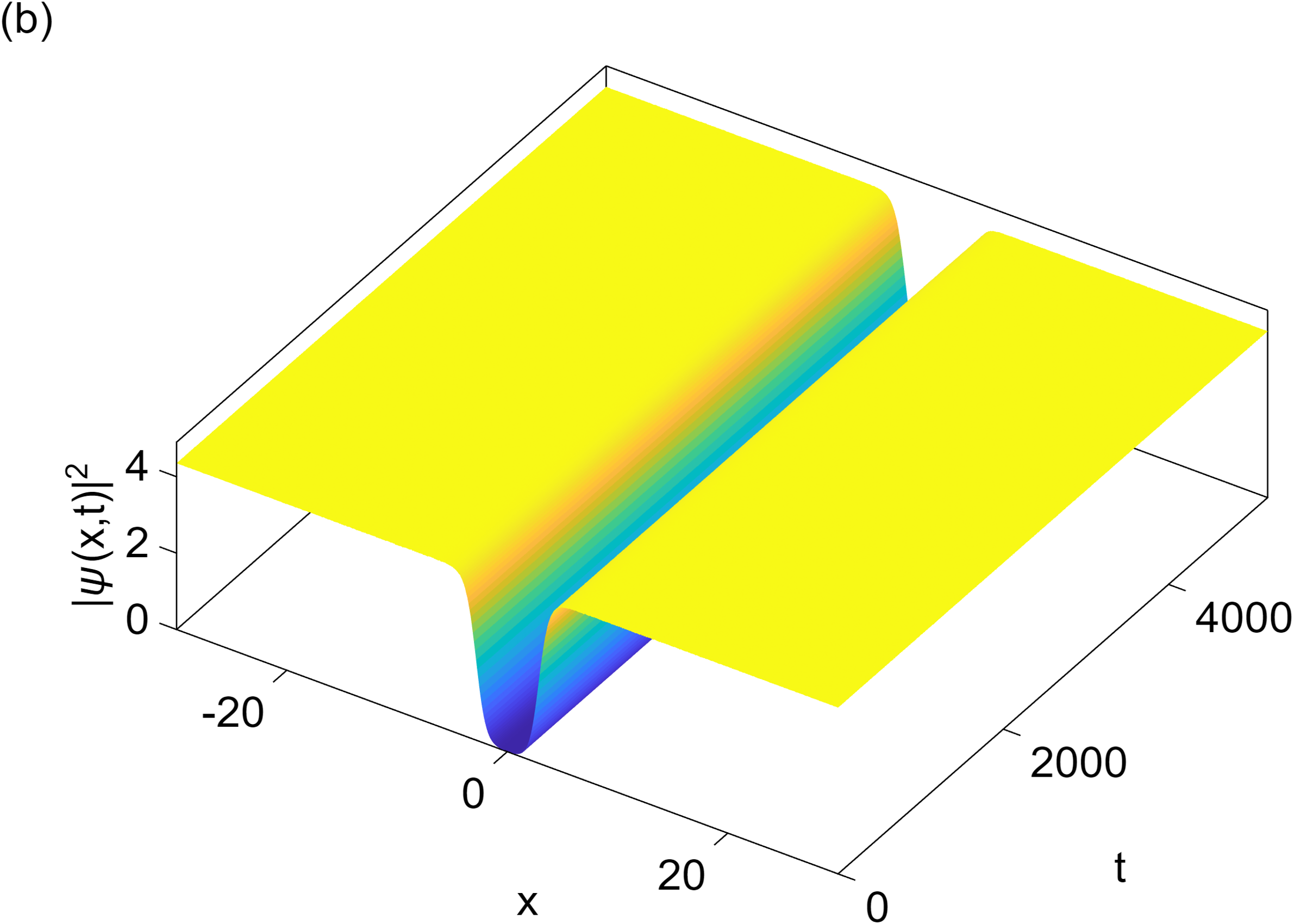}}
\caption{Stable evolution of slightly perturbed symmetric hole solutions with narrow and wide notch widths for $\mu=0.4$ (a) and $\mu=-0.72$ (b), respectively. Stability is tested by adding a random perturbation of amplitude $\epsilon=10^{-3}$. Other parameters are $\delta g=1$ and $g_{\mathrm{LHY}}=0.4$.}
\label{fig-dynSymHoles}
\end{figure}

The dynamical robustness of the symmetric holes is confirmed by numerical simulations. Figures~\ref{fig-dynSymHoles}(a) and \ref{fig-dynSymHoles}(b) depict the evolution of narrow and wide hole states, respectively, after a small random perturbation is added to the initial stationary profiles. The perturbation is added at the start of the simulations, and the later evolution is monitored. The absence of growth of the perturbation or breakup during evolution indicates that the symmetric hole solutions are dynamically stable. This result is consistent with the modulational stability of the upper constant-amplitude background supporting the holes.

\begin{figure}[htbp]
  \centerline{\includegraphics[width=8.5cm]{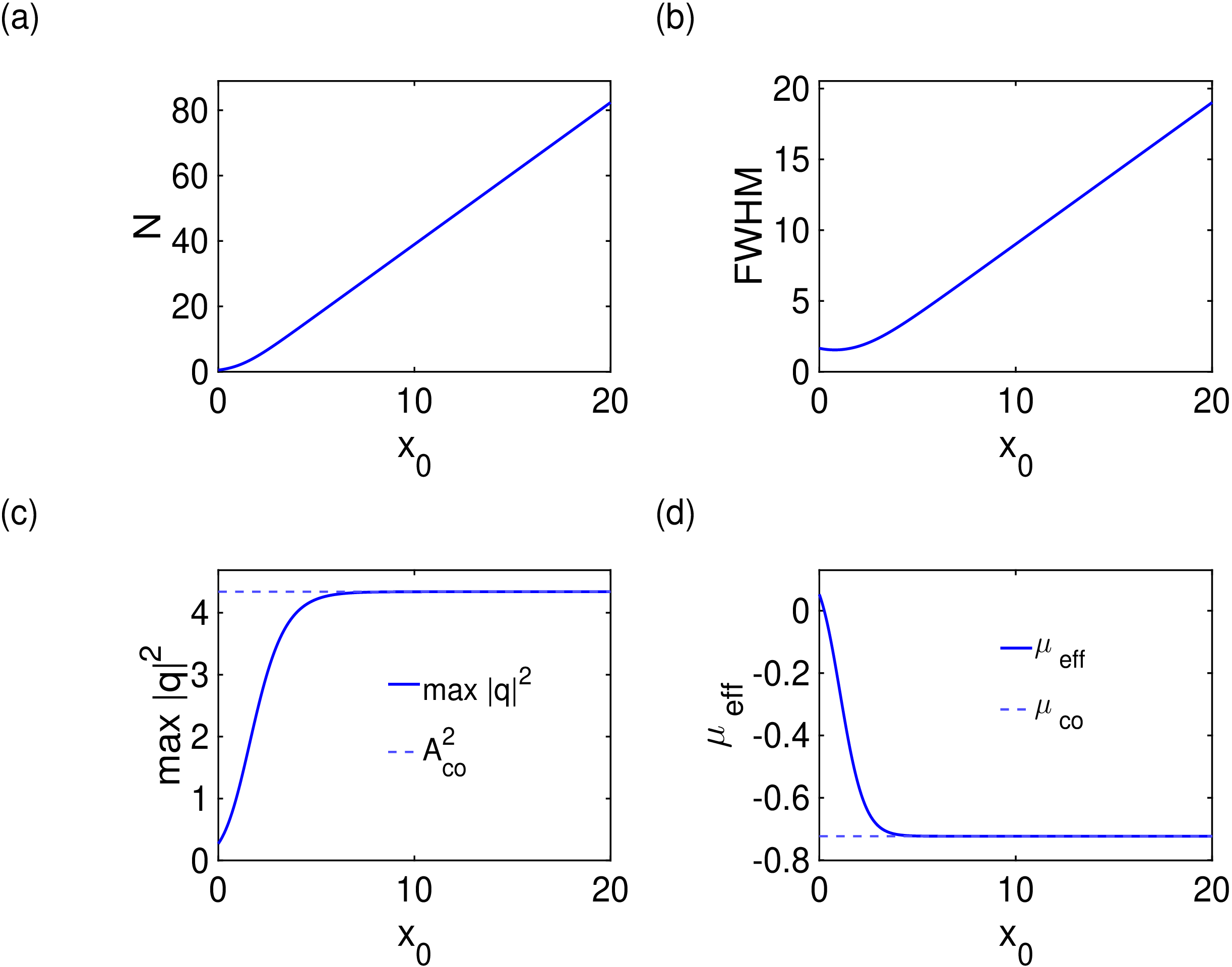}}
\caption{Dependence of quantum droplet parameters on separation distance $x_0$ for a kink-antikink superposition. 
(a) Particle number $N$ versus $x_0$.
(b) Full width at half maximum versus $x_0$.
(c) Peak density versus $x_0$, the horizontal dashed line indicates the coexistence value $A_{\mathrm{co}}^2$.
(d) Effective chemical potential $\mu_{\mathrm{eff}}$ versus $x_0$, the horizontal dashed line represents $\mu_{\mathrm{co}}$. 
The other parameters are $\delta g=1$ and $g_{\mathrm{LHY}}=0.4$.}
\label{fig-HolesParam}
\end{figure}

To clarify the relation between self-bound droplets and kink-type states, we construct approximate localized profiles from a kink-antikink pair. Let $q_{\mathrm{K}}(x)$ denote a kink solution connecting the vacuum state to the finite coexistence background, and let $q_{\mathrm{AK}}(x)$ denote the corresponding antikink. Their asymptotic values are defined as follows:
$$
q_{\mathrm{K}}(-\infty)=0, \qquad q_{\mathrm{K}}(+\infty)=A_{\mathrm{co}},
$$
and
$$
q_{\mathrm{AK}}(-\infty)=A_{\mathrm{co}}, \qquad q_{\mathrm{AK}}(+\infty)=0.
$$
A bright droplet-like localized state centered at $x=0$ can be constructed by placing the kink and antikink at $x=-x_0/2$ and $x=x_0/2$, respectively, using the multiplicative ansatz~(\ref{eq:kak_ansatz}):
\begin{equation}
\psi_{\mathrm{KAK}}(x,x_0)
=
\frac{
q_{\mathrm{K}}\left(x+\frac{x_0}{2}\right)
q_{\mathrm{AK}}\left(x-\frac{x_0}{2}\right)
}{
A_{\mathrm{co}}
}.
\label{eq:kak_ansatz}
\end{equation}
Division by $A_{\mathrm{co}}$ preserves the correct plateau amplitude. In the region between two well-separated fronts, both factors approach $A_{\mathrm{co}}$, so their product would give $A_{\mathrm{co}}^2$. Normalization by $A_{\mathrm{co}}$ ensures that the central flat-top part of the constructed state attains the physical coexistence amplitude $A_{\mathrm{co}}$ value. Outside the kink-antikink pair, at least one factor approaches zero, producing a spatially localized state. Thus, Eq.~(\ref{eq:kak_ansatz}) provides a natural interpolation between a bell-shaped localized profile at small separation and a flat-top droplet at large separation.

Figure~\ref{fig-HolesParam} summarizes how the main parameters of kink-antikink-based localized states depend on the separation distance $x_0$. The kink and antikink centers are positioned symmetrically with respect to the origin, so the two fronts are separated by a total distance of $x_0$. For small $x_0$, the two fronts strongly overlap, yielding a bell-shaped profile. As $x_0$ increases, the overlap diminishes, and a flat central region emerges, producing a flat-top droplet-like structure.

Figures~\ref{fig-HolesParam}(a) and \ref{fig-HolesParam}(b) indicate that both the particle number $N$ and the full width at half maximum increase with $x_0$. This trend is expected, as increasing the separation enlarges the nearly uniform central region between the kink and antikink. Figure~\ref{fig-HolesParam}(c) shows that the peak amplitude increases with $x_0$ and approaches the coexistence value $A_{\mathrm{co}}^2$. This saturation is a defining feature of a quantum droplet: once the central density reaches the equilibrium bulk value, additional particles primarily increase the droplet width rather than its peak density. Figure~\ref{fig-HolesParam}(d) presents the corresponding effective chemical potential $\mu_{\mathrm{eff}}$ as a function of $x_0$. As the separation increases, $\mu_{\mathrm{eff}}$ approaches the cutoff value $\mu_{\mathrm{co}}$, as indicated by the horizontal dashed line. This observation confirms that large kink-antikink separations correspond to the flat-top, liquid-like limit of the droplet family.

\begin{figure}[htbp]
  \centerline{\includegraphics[width=5.5cm]{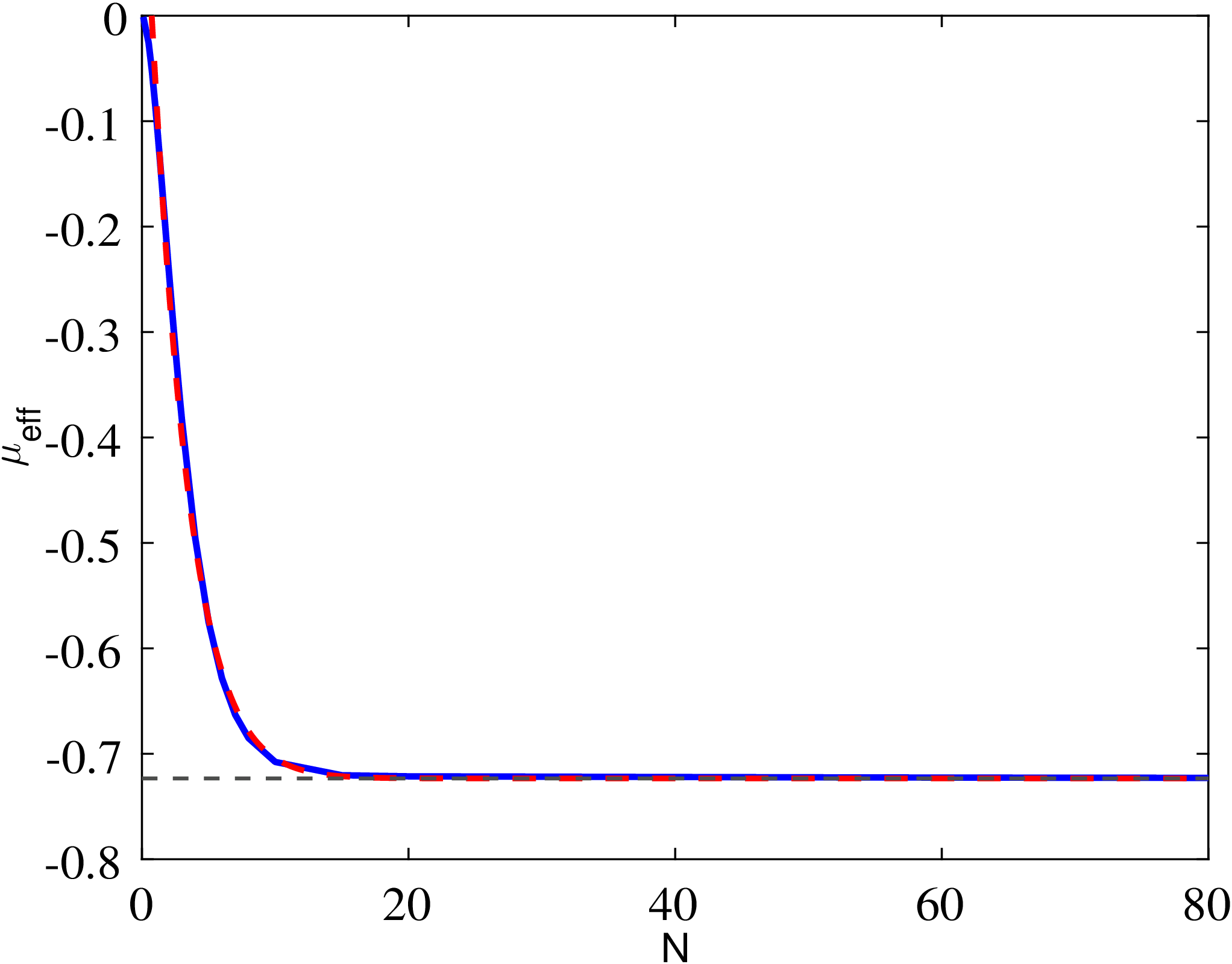}}
\caption{Chemical potential as a function of particle number. The red dashed line shows the effective chemical potential $\mu_{\mathrm{eff}}(N)$ for localized states constructed from a kink-antikink superposition with separation $x_0\in[0,20]$. The solid blue line shows $\mu(N)$ for self-bound quantum droplets obtained using the super-Gaussian trial function-based VA. The horizontal black dashed line marks the cutoff value $\mu_{\mathrm{co}}$. The parameters are $\delta g=1$ and $g_{\mathrm{LHY}}=0.4$.  }
\label{fig-HoleChemPot}
\end{figure}

Figure~\ref{fig-HoleChemPot} illustrates the relationship between the effective chemical potential and the particle number. The red dashed curve represents the kink-antikink construction, while the blue solid curve depicts the chemical potential of genuine self-bound quantum droplets obtained using the super-Gaussian trial function-based VA. The close agreement between these curves indicates that the kink-antikink ansatz captures the primary dependence $\mu(N)$ of the droplet family. Specifically, as $N$ increases, both approaches converge to the same limiting value $\mu_{\mathrm{co}}$. The negative slope of the curve, $d\mu/dN<0$, is consistent with the Vakhitov-Kolokolov stability condition for self-bound localized states\cite{VK}.

\begin{figure}[htbp]
  \centerline{\includegraphics[width=8.5cm]{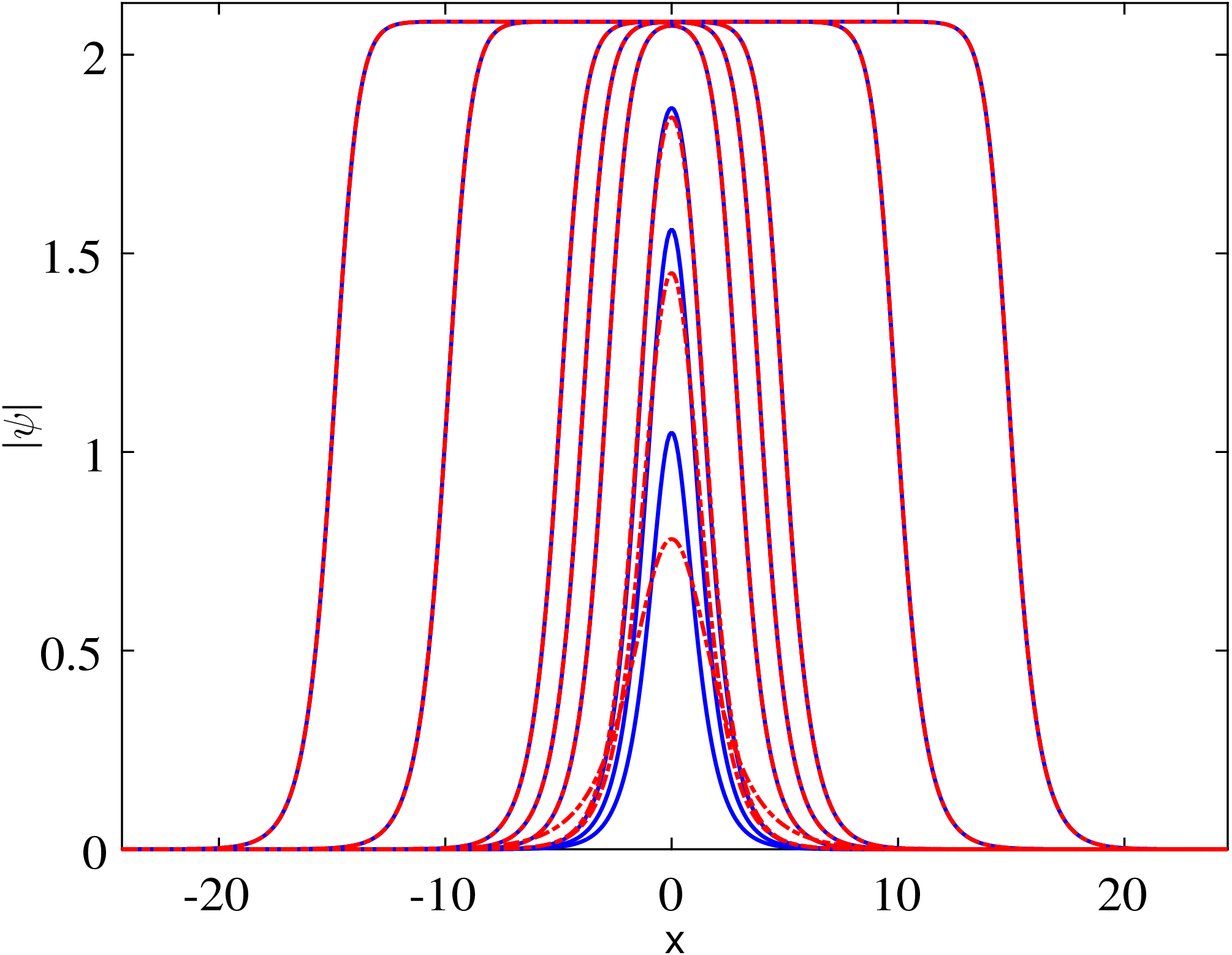}}
\caption{Profiles of localized states constructed from kink-antikink superpositions for different separation distances $x_0$. Blue solid lines correspond, from bottom to top, to $x_0=1,2,3,6,8,10,20,$ and $30$. For comparison, red dash-dotted lines show genuine self-bound quantum droplet profiles obtained by imaginary-time simulations at the same norms. The other parameters are $\delta g=1$ and $g_{\mathrm{LHY}}=0.4$.
}
\label{fig-HoleProfile}
\end{figure}

Figure~\ref{fig-HoleProfile} presents a direct comparison between the profiles generated by the kink-antikink ansatz and those of stationary self-bound quantum droplets obtained through imaginary-time simulations at identical norms. The blue solid curves correspond to the ansatz (\ref{eq:kak_ansatz}), while the red dash-dotted curves represent the numerically exact droplet profiles. For small separations, the agreement is qualitative but not exact, as the two fronts strongly overlap and the multiplicative ansatz does not fully reproduce the optimized stationary profile. As $x_0$ increases, the droplet becomes wider and flatter, the overlap region becomes less significant, and the agreement improves substantially. For large flat-top droplets, the kink-antikink construction closely reproduces the imaginary-time profiles. This result indicates that large self-bound droplets can be interpreted as localized states formed by a pair of well-separated nonlinear fronts.

\begin{figure}[htbp]
  \centerline{\includegraphics[width=4.3cm]{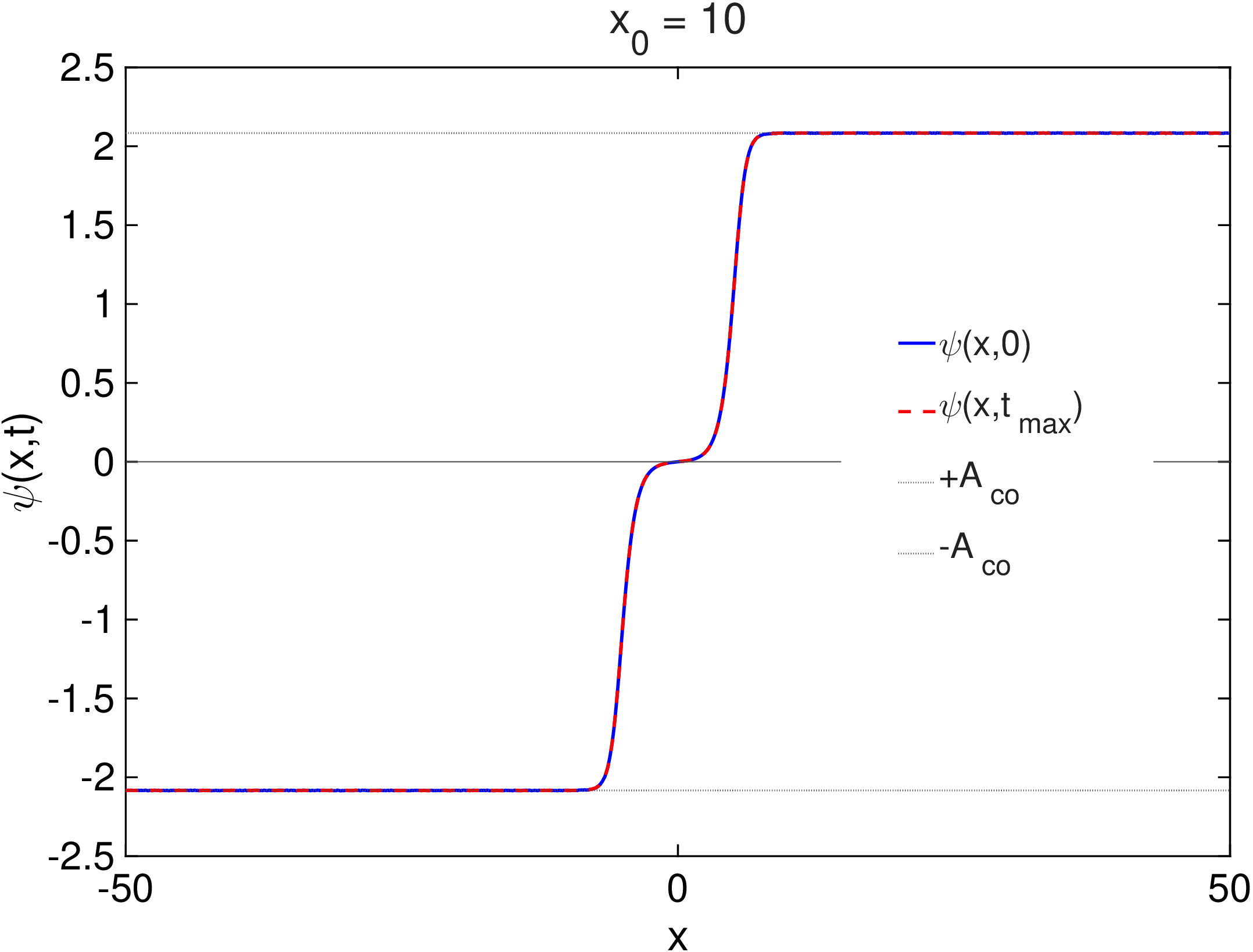}}
\caption{Initial and final profiles obtained by real-time propagation of an approximate dark-hole-like initial condition constructed from an antikink-kink pair with separation $x_0=10$. The solid blue curve shows the perturbed initial state at $t=0$, whereas the dashed red curve represents the state at $t=t_{\max}=10^4$. The horizontal dotted lines indicate the asymptotic backgrounds $\pm A_{\mathrm{co}}$. The initial condition satisfies $q_{\mathrm{dh}}(-\infty)=-A_{\mathrm{co}}$ and $q_{\mathrm{dh}}(+\infty)=+A_{\mathrm{co}}$, consequently, the sign reversal between the two outer backgrounds corresponds to a phase difference of $\pi$. The close agreement between the initial and final profiles shows that the density depletion and its phase structure persist throughout the simulated time interval. The initial state is perturbed multiplicatively by random noise of amplitude $\epsilon=10^{-3}$. The parameters are $\delta g=1$, $g_{\rm LHY}=0.4$.}
\label{fig-ProfAKKHole}
\end{figure}
A dark-hole-like state with a nontrivial phase difference is constructed by placing an antikink at $x=-x_0/2$ and a kink at $x=x_0/2$, with opposite signs. The resulting profile is

$$
q_{\mathrm{dh}}(x,x_0)=q_{\mathrm{K}}\left(x-\frac{x_0}{2}\right)-q_{\mathrm{AK}}\left(x+\frac{x_0}{2}\right).
$$

This profile satisfies $q_{\mathrm{dh}}(-\infty,x_0)=-A_{\mathrm{co}}$ and $q_{\mathrm{dh}}(+\infty,x_0)=+A_{\mathrm{co}}$. Consequently, the two asymptotic backgrounds possess equal densities but opposite signs, which corresponds to a phase difference of $\pi$. The density $|q_{\mathrm{dh}}|^2$ represents a localized depletion embedded within the finite background $A_{\mathrm{co}}^2$. The relative minus sign is crucial, as the plus-sign combination would approach the same phase on both sides and describe a bubble-like depletion without a phase jump.

The constituent fronts are exact stationary solutions at the coexistence chemical potential $\mu_{\mathrm{co}}$. However, their signed superposition at finite separation is an approximate initial condition rather than an exact stationary hole. Indeed, a finite-width stationary hole requires $a>A_{\mathrm{co}}$ and $\mu>\mu_{\mathrm{co}}$, whereas its width diverges at coexistence. We therefore use the constructed profile with $x_0=10$ to investigate the real-time persistence of a dark-hole-like configuration carrying a $\pi$ phase difference.

The real-time robustness of this state is examined by using
$
\psi(x,0)=q_{\mathrm{dh}}(x,x_0)\left[1+\epsilon R(x)\right],
$
where $\epsilon=10^{-3}$ and $R(x)$ is uniformly distributed over $[-1,1]$. Figure~\ref{fig-ProfAKKHole} compares the profiles $\psi(x,0)$ and $\psi(x,t_{\max})$ at $t_{\max}=10^4$. The two profiles are nearly indistinguishable. Specifically, the central depletion remains localized, the asymptotic amplitudes remain close to $\pm A_{\mathrm{co}}$, and the sign reversal associated with the $\pi$ phase difference is preserved. No growing instability is observed during the simulated interval, which demonstrates the finite-time dynamical robustness of the dark-hole-like state against the applied random perturbation.

	We next examine bubble-like states generated by an antikink-kink superposition. This construction directly connects isolated front solutions to finite-width density depletions embedded in the modulationally stable upper constant-amplitude background. The antikink-kink ordering produces a density depletion on a finite background, with the same phase in the two outer regions.

Let $q_{\mathrm{AK}}(x)$ denote a nonnegative antikink connecting the finite background to the vacuum, and let $q_{\mathrm{K}}(x)$ denote a nonnegative kink connecting the vacuum to the same finite background. Thus, $q_{\mathrm{AK}}(-\infty)=A_{\mathrm{co}}$, $q_{\mathrm{AK}}(+\infty)=0$, $q_{\mathrm{K}}(-\infty)=0$, and $q_{\mathrm{K}}(+\infty)=A_{\mathrm{co}}$. Placing the antikink at $x_{\mathrm{AK}}=-x_0/2$ and the kink at $x_{\mathrm{K}}=x_0/2$, the bubble-like profile is constructed as

$$
q_{\mathrm{AKK}}(x,x_0)=q_{\mathrm{AK}}\left(x+\frac{x_0}{2}\right)+q_{\mathrm{K}}\left(x-\frac{x_0}{2}\right).
$$

This profile approaches $+A_{\mathrm{co}}$ as $x\rightarrow\pm\infty$ and therefore carries no phase difference between its asymptotic backgrounds. For sufficiently large separation, it approximately follows the spatial ordering $A_{\mathrm{co}}\rightarrow 0\rightarrow A_{\mathrm{co}}$, producing a deep density depletion between two finite-density outer regions. For smaller $x_0$, the front tails overlap strongly, and the minimum density is appreciably nonzero. We refer to these profiles as dark-like and gray-like bubbles, respectively, according to their depletion depth. These terms describe the depletion depth and do not imply a phase jump. As in the signed construction, the finite-separation superposition of coexistence fronts is used as an approximate initial condition. The following simulations examine its finite-time dynamical robustness. 
To quantify the depletion, we use the normalized depth

$$
D(x_0)=1-\frac{\rho_{\min}(x_0)}{\rho_{\mathrm{co}}},
$$
$$
\rho_{\min}(x_0)=\min_x|q_{\mathrm{AKK}}(x,x_0)|^2, \qquad
\rho_{\mathrm{co}}=A_{\mathrm{co}}^2.
$$
With this definition, $D=0$ corresponds to no density depletion, whereas $D=1$ corresponds to a vanishing minimum density. In the numerical classification, we identify a profile as a dark-like bubble when $D(x_0)\geq D_{\mathrm{th}}$, with $D_{\mathrm{th}}=0.999$. Equivalently, this requires $\rho_{\min}(x_0)/\rho_{\mathrm{co}}\leq 10^{-3}$. Profiles under this threshold are classified as gray-like bubbles. For finite separation, the positive front tails generally leave a nonzero minimum density, even when this threshold is exceeded. This criterion diagnoses depletion depth and does not define a stability boundary.

\begin{figure}[htbp]
\centerline{\includegraphics[width=4.3cm]{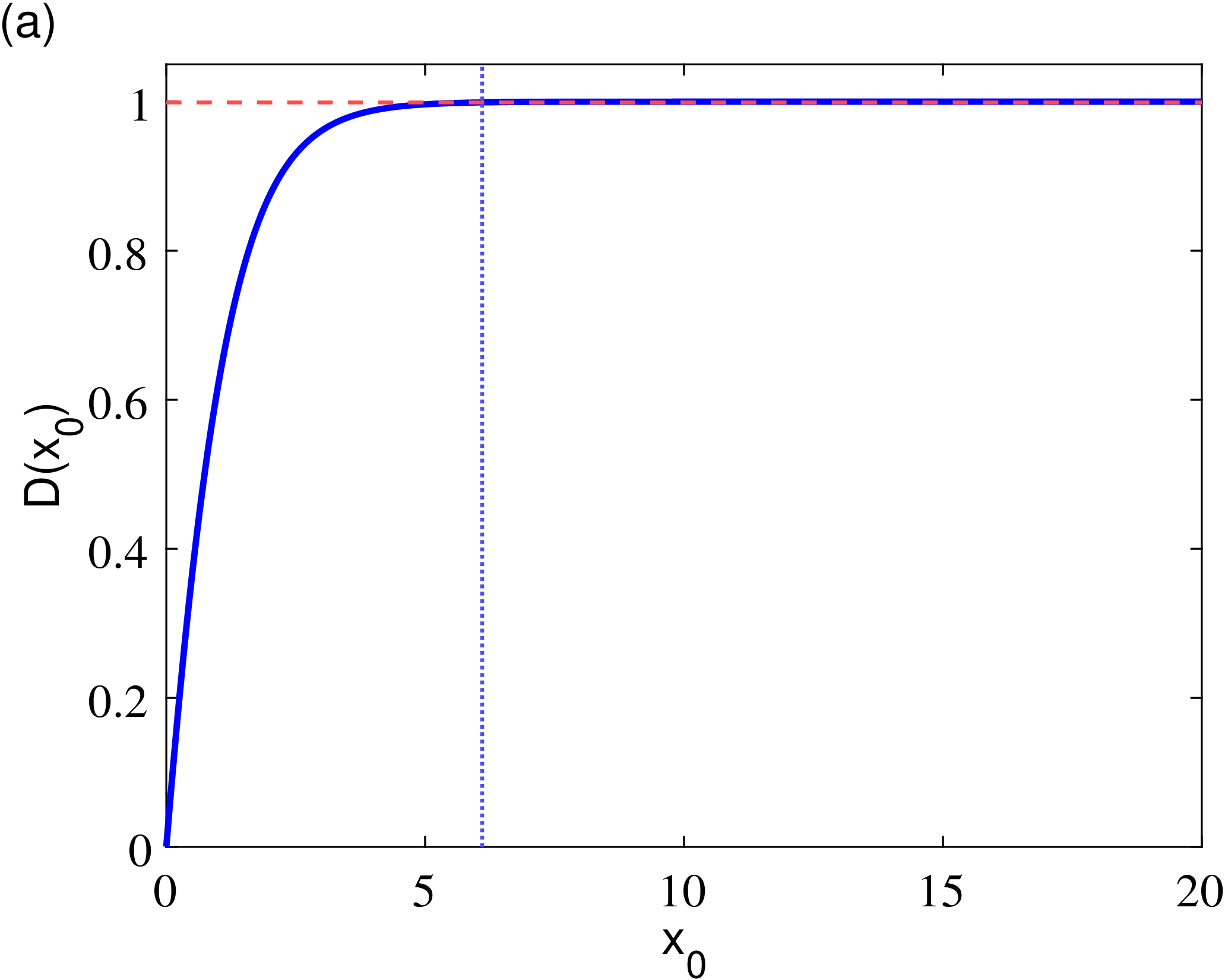}\includegraphics[width=4.9cm]{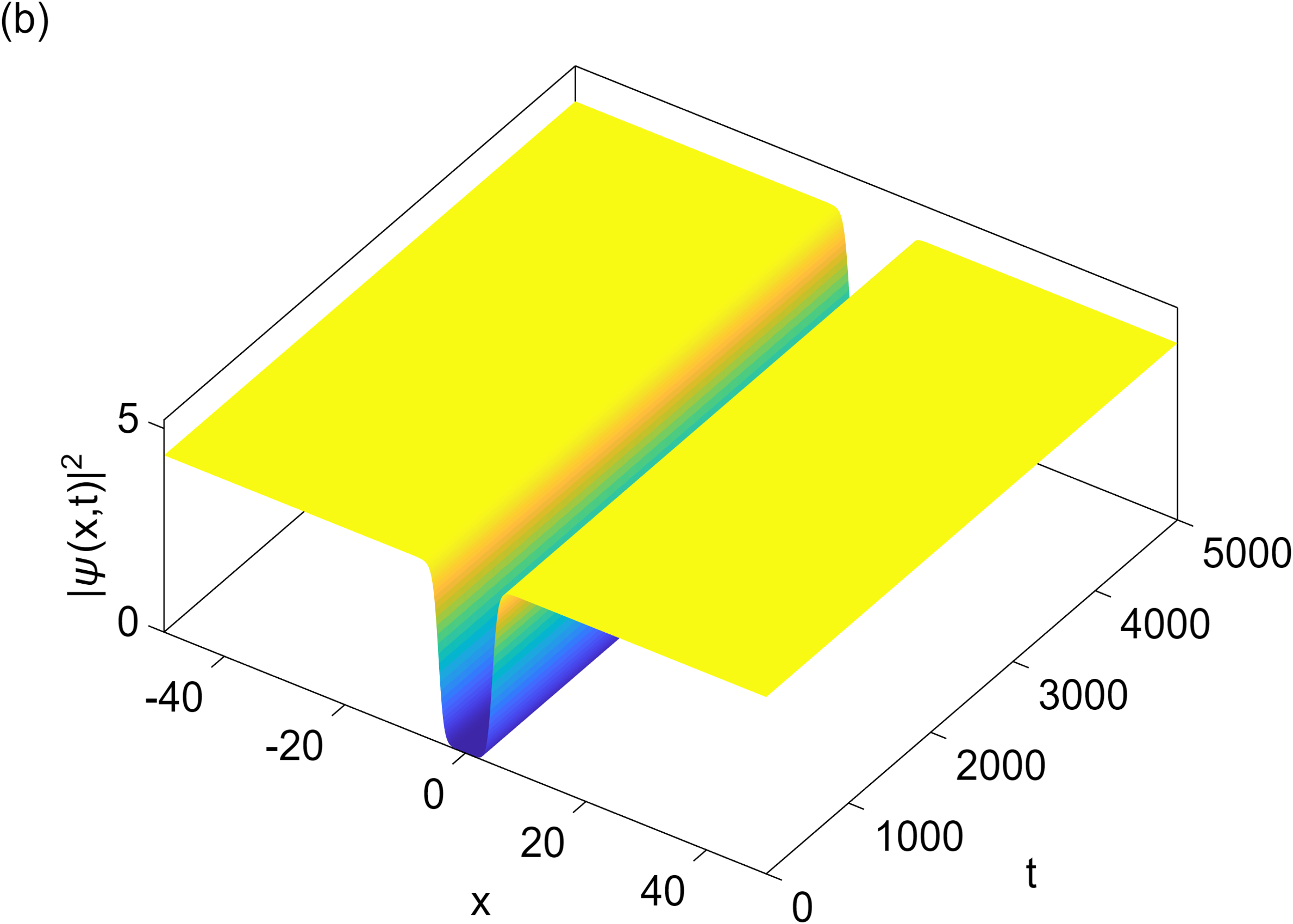}}
\caption{(a) Normalized depletion depth $D(x_0)$ versus the antikink-kink separation $x_0$. The horizontal red dashed line marks $D_{\mathrm{th}}=0.999$, and the vertical blue dotted line indicates the corresponding separation $x_0=6.1$. This threshold distinguishes gray-like and dark-like bubble profiles by depth. (b) Density evolution of a bubble-like initial state with $x_0=8$ under a small random perturbation, showing persistence of the depletion over the simulated time interval. The initial profile has no phase jump. The parameters are $\delta g=1$ and $g_{\mathrm{LHY}}=0.4$.}
\label{fig-DepthAndDynSymHoles}
\end{figure}

Figure~\ref{fig-DepthAndDynSymHoles}(a) shows that the depletion becomes deeper as the antikink-kink separation increases. The threshold $D_{\mathrm{th}}=0.999$ is reached at $x_0=6.1$. Below this separation, front-tail overlap produces an appreciable central density. Above it, the minimum density is sufficiently small for the profile to be classified as a dark-like bubble.
	Interpreting the dynamics requires distinguishing a deep depletion from a stable excitation. In the one-dimensional cubic-quadratic GPE, stationary bubbles without a phase jump are spectrally unstable throughout their existence interval~\cite{Katsimiga2023}. However, the Bogoliubov–de Gennes analysis reveals a decreasing instability growth rate toward the deep, wide-bubble limit, where the complementary atom number increases. In this regime, increasing the atom deficit lengthens the characteristic instability time, making the bubbles more persistent over finite observation times without rendering them spectrally stable~\cite{Katsimiga2023}. This decreasing-growth-rate trend applies near the deep-bubble limit rather than across the entire bubble family.
	Figure~\ref{fig-DepthAndDynSymHoles}(b) shows the real-time density evolution of the bubble-like initial state with $x_0=8$. The localized depletion continues under a small random perturbation without visible breakup over the simulated interval. This observation demonstrates finite-time dynamical robustness, rather than eigenvalue stability. The behavior is qualitatively consistent with the slow destabilization of deep bubbles reported in Ref.~\cite{Katsimiga2023}. However, because the present model contains cubic-quartic nonlinearities, establishing how the instability growth rate depends on $N_{\mathrm{def}}$ requires a separate Bogoliubov–de Gennes analysis of numerically converged stationary states. The modulational stability of the surrounding background alone does not guarantee stability of the localized bubble.

\section{Asymmetric two-component holes: homogeneous backgrounds, kink coexistence, and spinor induced instability}
\label{sec:AsymHoles}

We now extend the analysis to a two-component system in which the condensate components may have different chemical potentials, background amplitudes, and stationary profiles. Although the governing equations are invariant under the exchange of the two components, solutions with $\mu_1\neq\mu_2$ are generally asymmetric. This extension makes it possible to determine whether the existence and stability properties found in the scalar symmetric model persist when an additional relative-density, or spinor, degree of freedom is present.

The governing one-dimensional Gross-Pitaevskii system is
\begin{align}
i\frac{\partial \psi_j}{\partial t}
={}&
-\frac{1}{2}\frac{\partial^2\psi_j}{\partial x^2}
+\sigma\left(|\psi_j|^2-|\psi_{3-j}|^2\right)\psi_j
-\gamma|\psi_{3-j}|^2\psi_j
\notag\\
&+\delta_1
\left(|\psi_1|^2+|\psi_2|^2\right)^{3/2}\psi_j,
\qquad j=1,2 .
\label{eq:spinor_model}
\end{align}
Here $\gamma>0$ describes the attractive intercomponent mean-field interaction, $\delta_1>0$ is the coefficient of the repulsive Beyond mean-field contribution, and $\sigma$ controls the energetic cost of an imbalance between the two component densities. Unless stated otherwise, the parameters are $\sigma=10$, $\gamma=1$, $\delta=0.4$, and $\delta_1=\delta/2^{3/2}$.

The energy functional is
\begin{equation*}
 E=\int dx\left[
 \frac{1}{2}\sum_{j=1}^2|\partial_x\psi_j|^2
 +\mathcal{E}(n_1,n_2)
 \right],
\end{equation*}
where $n_j=|\psi_j|^2$, $\rho=n_1+n_2$, $d=n_1-n_2$, $C=2\sigma+\gamma$, and
\begin{eqnarray*}
 &\mathcal{E}(n_1,n_2)
 =\frac{\sigma}{2}(n_1-n_2)^2-\gamma n_1n_2
 +\frac{2\delta_1}{5}\rho^{5/2} \\
\nonumber 
 &=-\frac{\gamma}{4}\rho^2+\frac{C}{4}d^2
 +\frac{2\delta_1}{5}\rho^{5/2}.
\end{eqnarray*}
The component chemical potentials are $\mu_j=\partial\mathcal{E}/\partial n_j$.

\subsection{Homogeneous backgrounds and branch structure}
\label{sec:AsymMI}

We first consider homogeneous solutions of the form $\psi_j(x,t)=a_j e^{-i\mu_jt}$, where $a_j\geq0$. Introducing the component densities $n_j=a_j^2$, the total density $\rho=n_1+n_2$, and the density imbalance $d=n_1-n_2$, substitution into Eq.~(\ref{eq:spinor_model}) yields
$$
\mu_1
=
\sigma n_1-(\sigma+\gamma)n_2+\delta_1\rho^{3/2},
\,\,
\mu_2
=
\sigma n_2-(\sigma+\gamma)n_1+\delta_1\rho^{3/2}.
$$
The difference between these relations determines the component imbalance as
$$
d=\frac{\mu_1-\mu_2}{C},
\qquad
C=2\sigma+\gamma,
$$
whereas their average, $\bar{\mu}=(\mu_1+\mu_2)/2$, depends only on the total density:
$$
\bar{\mu}
=
-\frac{\gamma}{2}\rho+\delta_1\rho^{3/2}.
$$
Thus, the homogeneous problem separates naturally into a total density channel, described by $\rho$, and a relative-density channel, described by $d$.

Upon introducing $y=\sqrt{\rho}$, the equation for the total density becomes the cubic algebraic equation
$$
\delta_1y^3-\frac{\gamma}{2}y^2-\bar{\mu}=0.
$$
Only positive real roots satisfying $\rho\geq |d|$ are physically admissible, because this condition guarantees that both component densities remain nonnegative. Once an admissible root has been selected, the individual densities are
$$
n_1=\frac{\rho+d}{2},
\qquad
n_2=\frac{\rho-d}{2},
$$
or, equivalently,
$$
a_1^2
=
\frac{1}{2}
\left[
\rho+\frac{\mu_1-\mu_2}{2\sigma+\gamma}
\right],
\qquad
a_2^2
=
\frac{1}{2}
\left[
\rho-\frac{\mu_1-\mu_2}{2\sigma+\gamma}
\right].
$$

The turning point of the homogeneous family follows from $d\bar{\mu}/d\rho=0$. It is located at
$$
\rho_{\mathrm{cr}}
=
\frac{\gamma^2}{9\delta_1^2},
\qquad
\bar{\mu}_{\mathrm{cr}}
=
-\frac{\gamma^3}{54\delta_1^2}.
$$
For $\bar{\mu}_{\mathrm{cr}}<\bar{\mu}<0$, the cubic equation has two positive roots. The solution with $\rho_+>\rho_{\mathrm{cr}}$ defines the upper branch, while the root satisfying $0<\rho_-<\rho_{\mathrm{cr}}$ defines the lower branch. The two branches merge at $\bar{\mu}=\bar{\mu}_{\mathrm{cr}}$. For $\bar{\mu}>0$, only the upper positive-density branch remains.

In the $(\mu_1,\mu_2)$ plane, the branch-merging condition is
$$
\frac{\mu_1+\mu_2}{2}
=
-\frac{\gamma^3}{54\delta_1^2}.
$$
Consequently, when $\mu_2$ is fixed, the critical chemical potential of the first component is
$$
\mu_{1,\mathrm{cr}}(\mu_2)
=
-\frac{\gamma^3}{27\delta_1^2}-\mu_2,
$$
provided that the corresponding values of $n_1$ and $n_2$ satisfy the positivity condition.

\subsection{Vacuum coexistence and the kink limit}
\label{sec:AsymKAK}

The kink limit is distinct from the branch-merging point. The latter indicates where the two homogeneous-density branches coalesce, whereas the kink limit corresponds to coexistence between a finite-density homogeneous phase and the vacuum.

For a homogeneous two-component state, the nonlinear energy density associated with Eq.~(\ref{eq:spinor_model}) is
$$
\mathcal{E}(\rho,d)
=
-\frac{\gamma}{4}\rho^2
+\frac{C}{4}d^2
+\frac{2\delta_1}{5}\rho^{5/2}.
$$
For fixed component chemical potentials, the grand potential functional is
$$
\Omega
=
E-\mu_1N_1-\mu_2N_2,
\qquad
N_j=\int n_j\,dx.
$$
For a homogeneous state, the grand potential density is
$$
\omega(\rho,d;\mu_1,\mu_2)
=
\mathcal{E}(\rho,d)-\mu_1n_1-\mu_2n_2.
$$
The thermodynamic pressure is the negative grand potential density,
$$
p
=
-\omega
=
\mu_1n_1+\mu_2n_2-\mathcal{E}
=
-\frac{\gamma}{4}\rho^2
+\frac{C}{4}d^2
+\frac{3\delta_1}{5}\rho^{5/2}.
$$
The same expression follows from
$p=\sum_jn_j(\partial\mathcal{E}/\partial n_j)-\mathcal{E}$,
or from $p=-(\partial E/\partial L)_{N_1,N_2}$ for a uniform system. Equivalently, $-p$ is the grand potential density $\mathcal{E}-\mu_1n_1-\mu_2n_2$. For the vacuum, $n_1=n_2=0$, and both the energy density, grand potential density, and pressure vanish. Therefore, coexistence between the vacuum and a finite-density phase requires the pressure of the latter to vanish as well.

This condition has a direct interpretation for a stationary kink. A kink is a front whose amplitudes approach the vacuum on one side and a nonzero homogeneous state on the other. In the stationary spatial problem, these two asymptotic states must have the same value of the first integral, or equivalently the same grand potential density. If their pressures were different, one phase would be thermodynamically favoured and the interface would move, expand, or contract rather than remain stationary. Thus, the zero-pressure condition ensures that neither the vacuum nor the finite-density phase drives the front into the other.

Using the homogeneous chemical potentials, the pressure takes the form given above. We introduce the normalized density imbalance
$$
P
=
\frac{d}{\rho}
=
\frac{n_1-n_2}{n_1+n_2},
\qquad |P|\leq1.
$$
The parameter $P$ may also be interpreted as a polarization parameter: $P=0$ corresponds to a balanced mixture, while $P=1$ and $P=-1$ describe complete occupation of the first and second components, respectively.

The zero-pressure condition gives the coexistence density
$$
\rho_{\mathrm{co}}(P)
=
\left[
\frac{5}{12\delta_1}
\left(\gamma-CP^2\right)
\right]^2.
$$
A finite coexistence density requires $\gamma-CP^2>0$. In addition, the background connected to the vacuum by the kink must belong to the upper branch. The condition $\rho_{\mathrm{co}}>\rho_{\mathrm{cr}}$ is equivalent to
$$
P^2<\frac{\gamma}{5C}.
$$
This condition is more restrictive than the mere positivity of $\rho_{\mathrm{co}}$ and selects the relevant coexistence root.

The average chemical potential at coexistence is
$$
\bar{\mu}_{\mathrm{co}}(P)
=
-\frac{25}{1728\delta_1^2}
\left(\gamma-CP^2\right)^2
\left(\gamma+5CP^2\right),
$$
while the individual chemical potentials are
$$
\mu_{1,\mathrm{co}}
=
\bar{\mu}_{\mathrm{co}}
+\frac{1}{2}CP\rho_{\mathrm{co}},
\qquad
\mu_{2,\mathrm{co}}
=
\bar{\mu}_{\mathrm{co}}
-\frac{1}{2}CP\rho_{\mathrm{co}}.
$$
For fixed $\mu_2$, the coexistence polarization $P_{\mathrm{co}}$ is obtained from $\mu_{2,\mathrm{co}}(P_{\mathrm{co}})=\mu_2$. The corresponding $\mu_{1,\mathrm{co}}$ then follows from the first relation. If several algebraic roots are found, only the root satisfying $\rho_{\mathrm{co}}>\rho_{\mathrm{cr}}$ is retained.

For $\mu_2=-0.5$ and the parameters used in the numerical calculations, the physical root is $P_{\mathrm{co}}\simeq-2.45\times10^{-3}$, yielding $\rho_{\mathrm{co}}\simeq8.67836$ and $\mu_{1,\mathrm{co}}\simeq-0.94731$. The small value of $|P_{\mathrm{co}}|$ shows that the two components become nearly balanced close to the kink limit, even though their chemical potentials remain different.

For the exactly symmetric state, $P=0$, one obtains
$$
\rho_{\mathrm{co}}
=
\left(\frac{5\gamma}{12\delta_1}\right)^2,
\qquad
\mu_{1,\mathrm{co}}
=
\mu_{2,\mathrm{co}}
=
-\frac{25\gamma^3}{1728\delta_1^2}.
$$
This result reduces precisely to the coexistence condition obtained for the scalar symmetric model. Indeed, when $\psi_1=\psi_2=\psi$, the imbalance term proportional to $\sigma$ vanishes and, because $\delta_1=\delta/2^{3/2}$, Eq.~(\ref{eq:spinor_model}) reduces to
$$
i\psi_t
=
-\frac{1}{2}\psi_{xx}
-\gamma|\psi|^2\psi
+\delta|\psi|^3\psi.
$$
The per-component coexistence amplitude and chemical potential are then
$$
A_{\mathrm{co}}
=
\frac{5\gamma}{6\delta},
\qquad
\mu_{\mathrm{co}}
=
-\frac{25\gamma^3}{216\delta^2},
$$
which coincide with the results of the preceding symmetric analysis after identifying $\gamma$ with the attractive cubic coefficient and $\delta$ with the repulsive LHY coefficient.

\subsection{Modulational spectra of the homogeneous states}
\label{sec:AsymMISepectr}
The stability of a homogeneous background is examined by introducing Bogoliubov perturbations,
$$
\psi_j = \left[ a_j + U_j e^{i(kx-\omega t)} + V_j^* e^{-i(kx-\omega^*t)}
\right] e^{-i\mu_jt}.
$$
Here $U_j$ and $V_j$ are the small amplitudes of the positive- and negative-wavenumber sidebands in the $j$th component. Their coupling accounts for the simultaneous modulation of the amplitude and phase of each condensate component.

Linearization around the constant background leads to the nonlinear stiffness matrix
$$
\mathsf{B}
=
\begin{pmatrix}
n_1h_d & \sqrt{n_1n_2}\,h_o\\
\sqrt{n_1n_2}\,h_o & n_2h_d
\end{pmatrix},
$$
where
$$
h_d
=
\sigma+\frac{3}{2}\delta_1\sqrt{\rho},
\qquad
h_o
=
-(\sigma+\gamma)+\frac{3}{2}\delta_1\sqrt{\rho}.
$$
The diagonal coefficient $h_d$ describes the response of each component to its own density variation, whereas $h_o$ describes the response induced by a density variation in the other component. The matrix $\mathsf{B}$ therefore determines whether small coupled density perturbations experience a restoring or an anti-restoring nonlinear response.

Its eigenvectors describe two collective perturbation channels. In the balanced limit, these reduce to an in-phase total density mode and an out-of-phase relative-density, or polarization, mode. For an asymmetric background, the two channels are generally mixed. The corresponding stiffness eigenvalues are
$$
\lambda_\pm
=
\frac{1}{2}
\left[
h_d\rho
\pm
\sqrt{
h_d^2d^2+h_o^2(\rho^2-d^2)
}
\right].
$$
They determine the two Bogoliubov branches,
$$
\omega_\pm(k)
=
\frac{k}{2}
\sqrt{k^2+4\lambda_\pm},
\qquad k\geq0.
$$
At long wavelengths, $\omega_\pm\simeq k\sqrt{\lambda_\pm}$, so that $\lambda_\pm$ represent the squared sound velocities of the two collective modes. A negative value of $\lambda_-$ produces an imaginary sound velocity and hence modulational instability.

When $\lambda_-<0$, the unstable band is $0<k<2\sqrt{-\lambda_-}$, and the corresponding growth rate is
$$
\Gamma(k)
=
\operatorname{Im}\omega_-(k)
=
\frac{k}{2}\sqrt{-k^2-4\lambda_-}.
$$
The relation between the branch structure and modulational stability follows from
$$
\det\mathsf{B}
=
n_1n_2\,C
\left(
-\gamma+3\delta_1\sqrt{\rho}
\right).
$$
Since the trace of $\mathsf{B}$ is $h_d\rho>0$, the sign of the determinant determines whether the smaller eigenvalue is positive or negative. On the upper branch, $\rho>\rho_{\mathrm{cr}}$, so $\det\mathsf{B}>0$ and both eigenvalues are nonnegative. The upper homogeneous background is therefore modulationally stable. On the lower branch, $\rho<\rho_{\mathrm{cr}}$, the determinant is negative and $\lambda_-<0$, producing a long-wavelength modulational instability. At the branch-merging point, $\lambda_-=0$.

For a balanced state, the two stiffnesses simplify to
$$
\lambda_{\mathrm{dens}}
=
\frac{\rho}{2}
\left(
-\gamma+3\delta_1\sqrt{\rho}
\right),
\qquad
\lambda_{\mathrm{pol}}
=
\frac{C\rho}{2}.
$$
The polarization channel remains positive, while the density channel changes sign at $\rho=\rho_{\mathrm{cr}}$. This makes clear that the modulational instability of the lower branch is driven by the total density mode rather than by the relative-density mode.

\subsection{Asymmetric holes and spinor induced instability}
\label{sec:AsymSpinorHoles}
We next construct stationary sign changing holes in the form $\psi_j(x,t)=q_j(x)e^{-i\mu_jt}$. The real profiles satisfy $q_j(-\infty)=-a_j$ and $q_j(+\infty)=a_j$, and both components pass through zero at the center of the notch. The asymptotic amplitudes are selected from the modulationally stable upper branch.

The total density deficit is defined as
$$
N
=
\int_{-\infty}^{+\infty}
\left[
a_1^2-|q_1|^2
+
a_2^2-|q_2|^2
\right]dx,
$$
and the integral notch width is
$$
w
=
2
\left[
\frac{
\displaystyle
\int_{-\infty}^{+\infty}
x^2
\left[
a_1^2-|q_1|^2
+
a_2^2-|q_2|^2
\right]dx
}{
N
}
\right]^{1/2}.
$$
The quantity $N$ measures the total number of particles removed from the two homogeneous backgrounds by the notch, while $w$ characterizes the spatial extent of this density deficit.

\begin{figure}[htbp]
  \centerline{\includegraphics[width=4.45cm]{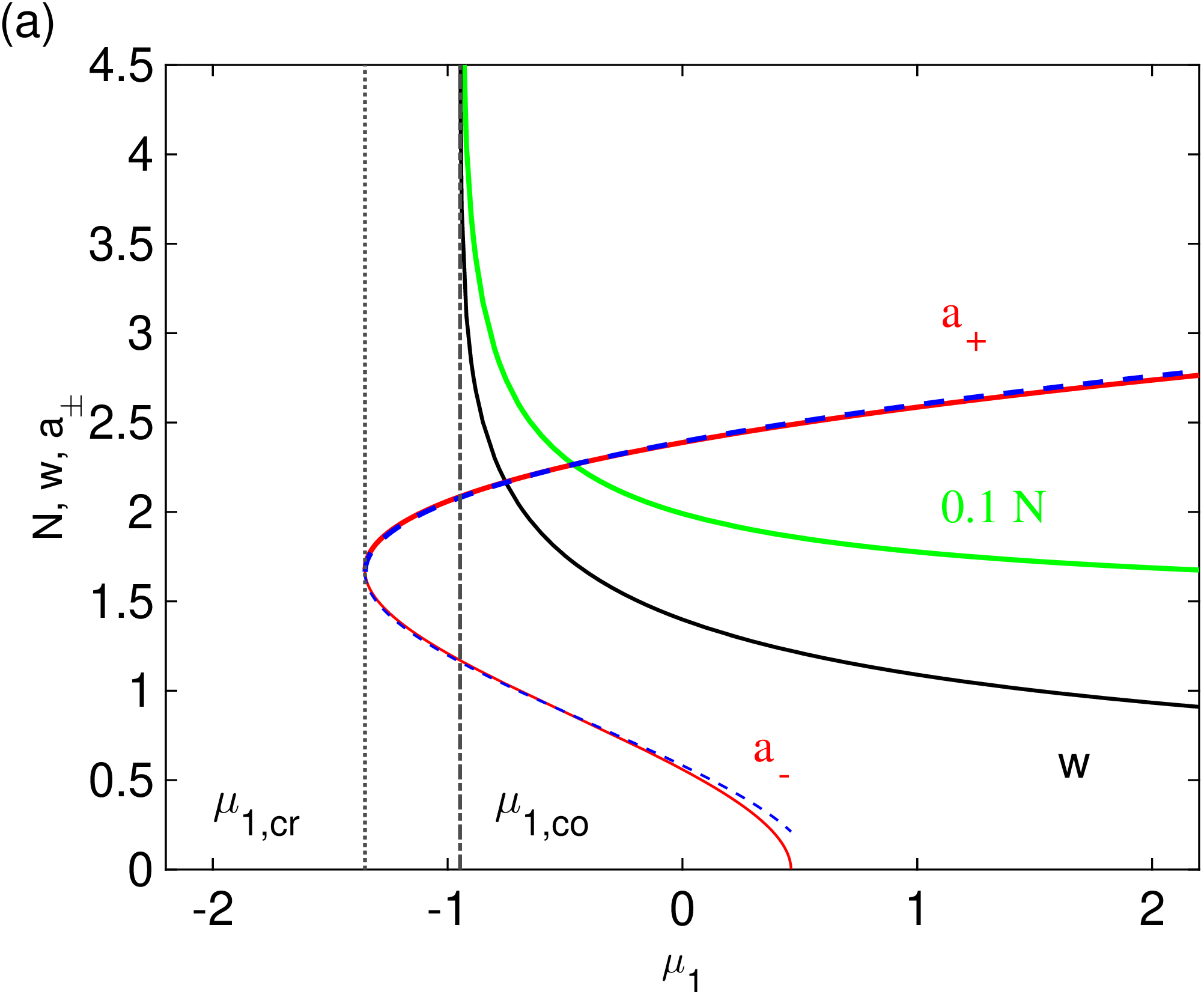}
  \includegraphics[width=4.45cm]{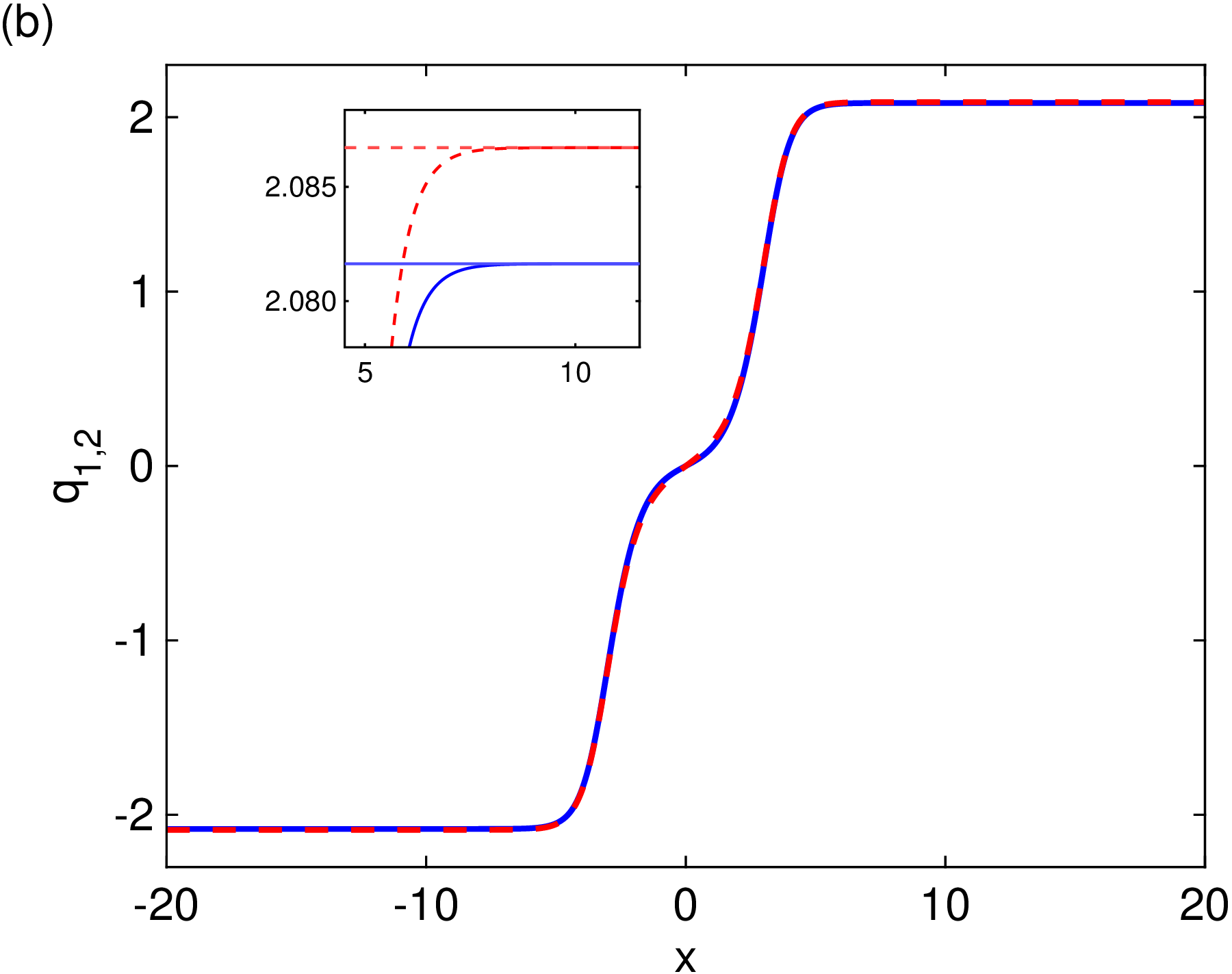}}
  \centerline{\includegraphics[width=4.45cm]{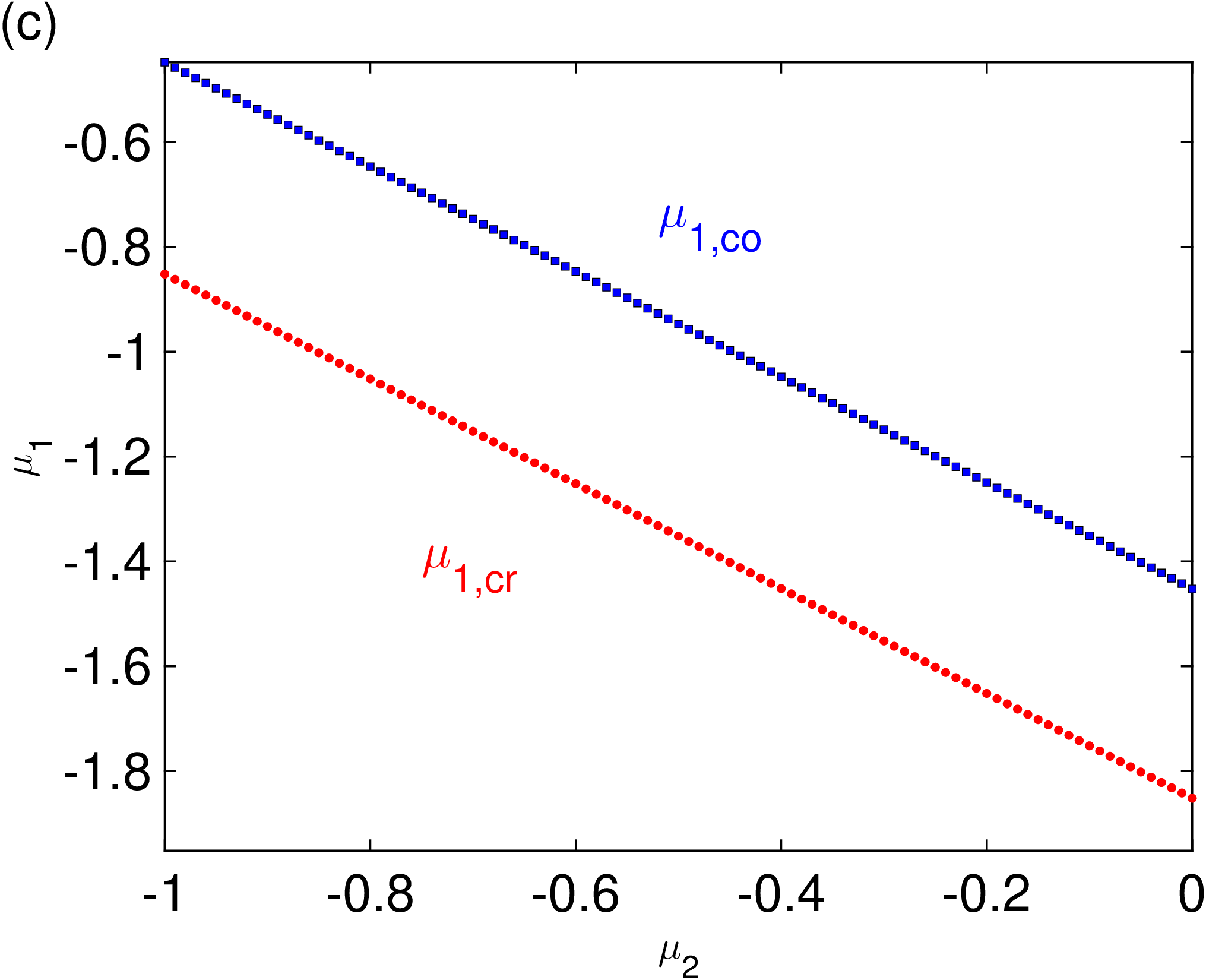}}
\caption{(a) Parameters of asymmetric holes as functions of the chemical potential $\mu$. The left and right vertical dotted lines indicate $\mu_{1, \mathrm{cr}}$ and $\mu_{1,\mathrm{co}}$, respectively.
The green curve represents the redefined norm $N(\mu)$, scaled by $0.1$ for clarity, and the black curve depicts the notch width $w(\mu)$.
The blue dashed and red curves denote the constant-background branches $a_{1,\pm}(\mu_1)$ and $a_{2,\pm}(\mu_1)$, respectively, with thick and thin segments corresponding to the upper and lower branches.
(b) Representative profiles of the asymmetric holes for $\mu_{1}=-0.945$ (red dashed line) and $\mu_2=-0.5$ (blue solid line).
The inset highlights the region of the main plot that demonstrates the distinction between the profiles. The horizontal red dashed and solid blue lines indicate the stable upper branches of the constant-amplitude solutions for the respective components.
(c) Dependencies of the first component's critical $\mu_{1,\mathrm{cr}}$ and cutoff $\mu_{1,\mathrm{co}}$ chemical potential values on the second component chemical potential $\mu_2$. The remaining parameters are $\sigma=10$, $\gamma=1$, $\delta=0.4$, and $\delta_1=\delta/2^{3/2}$.}
\label{fig-kk63}
\end{figure}

Figure~\ref{fig-kk63}(a) summarizes the asymmetric-hole family obtained by varying $\mu_1$ at fixed $\mu_2=-0.5$. The green curve shows the redefined norm $N$, multiplied by $0.1$ for visual comparison with the other quantities, whereas the black curve represents the notch width $w$. Both curves grow strongly as $\mu_1$ approaches $\mu_{1,\mathrm{co}}$ from above. This increase reflects the progressive separation of the two fronts forming the hole and the development of a broad, nearly empty central region.

The blue dashed curves in Fig.~\ref{fig-kk63}(a) represent the first-component background amplitudes $a_{1,\pm}$, while the red curves represent $a_{2,\pm}$. The thick parts correspond to the upper, modulationally stable branches, and the thin parts indicate the lower, modulationally unstable branches. Because $\mu_1\neq\mu_2$, the blue and red curves do not coincide, demonstrating the finite population imbalance between the two components.

The left vertical dotted line in Fig.~\ref{fig-kk63}(a) marks $\mu_{1,\mathrm{cr}}$, where the upper and lower constant-background branches merge. The right vertical dotted line marks $\mu_{1,\mathrm{co}}$, where the hole broadens into two asymptotically separated kink fronts. The finite interval between these two lines shows that the branch-merging and kink coexistence thresholds are physically distinct: constant-amplitude solutions already exist above $\mu_{1,\mathrm{cr}}$, but the hole family is bounded by the higher value $\mu_{1,\mathrm{co}}$.

Figure~\ref{fig-kk63}(b) compares the two components of a representative asymmetric hole with $\mu_1=-0.945$ and $\mu_2=-0.5$. The red dashed curve represents the first-component profile $q_1$, while the blue solid curve represents $q_2$. Their unequal asymptotic amplitudes and slightly different transition shapes arise from the component imbalance. The horizontal red dashed and blue solid lines show the corresponding upper-branch background amplitudes. The inset enlarges the region in which the two profiles differ most visibly. Despite these differences, both components form a common sign changing notch and approach their respective stable backgrounds away from its center.

Figure~\ref{fig-kk63}(c) compares the critical and cutoff chemical potentials of the first component as functions of $\mu_2$. The critical curve $\mu_{1,\mathrm{cr}}$ follows the exact linear relation derived from the branch-merging condition. In contrast, the cutoff curve $\mu_{1,\mathrm{co}}$ is obtained parametrically from the zero-pressure coexistence condition and is generally nonlinear. The cutoff curve lies above the critical curve throughout the displayed range. Their separation identifies the range in which homogeneous backgrounds exist but the kink-connected hole family has not yet emerged.

\begin{figure}[htbp]
  \centerline{\includegraphics[width=4.45cm]{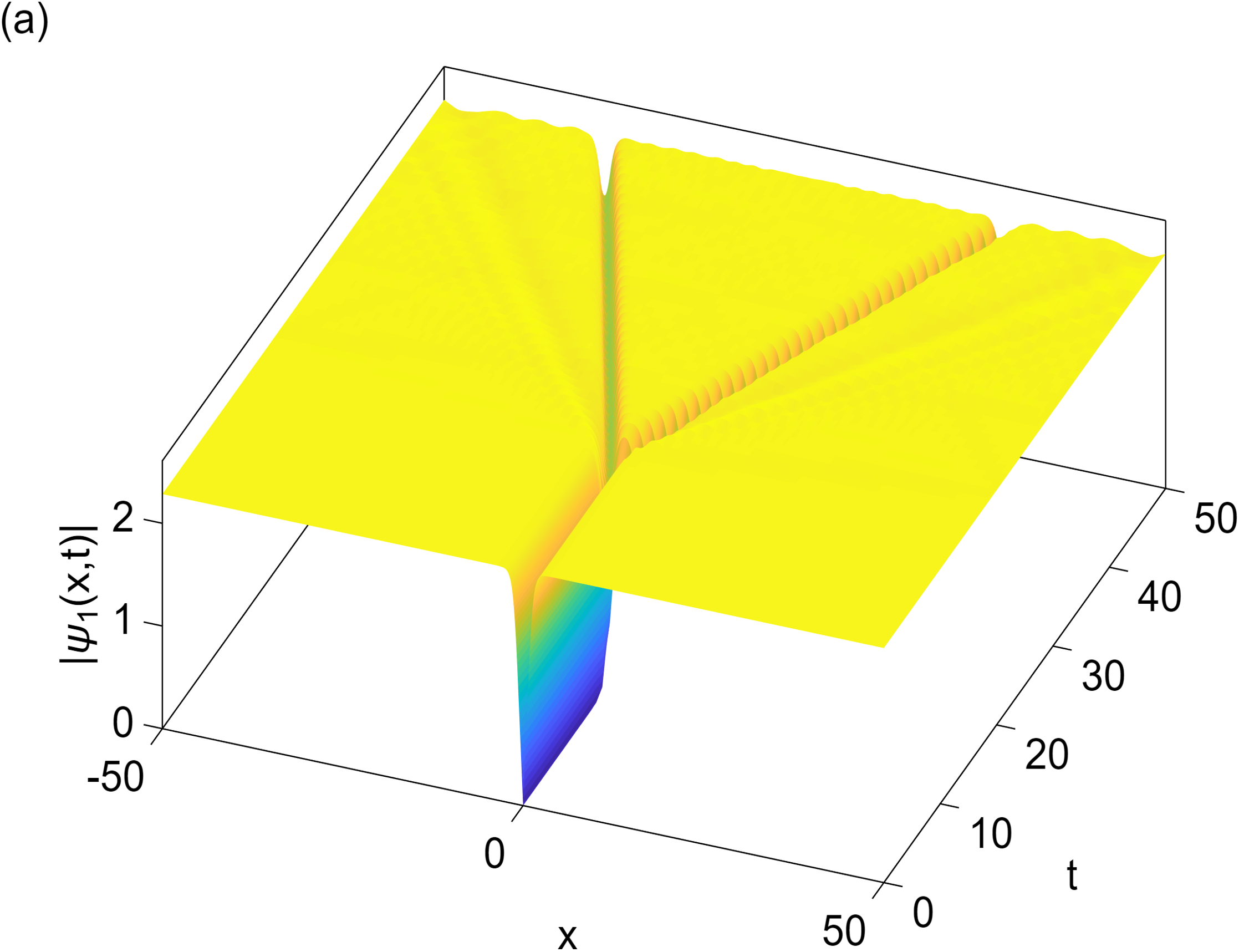}
  \includegraphics[width=4.45cm]{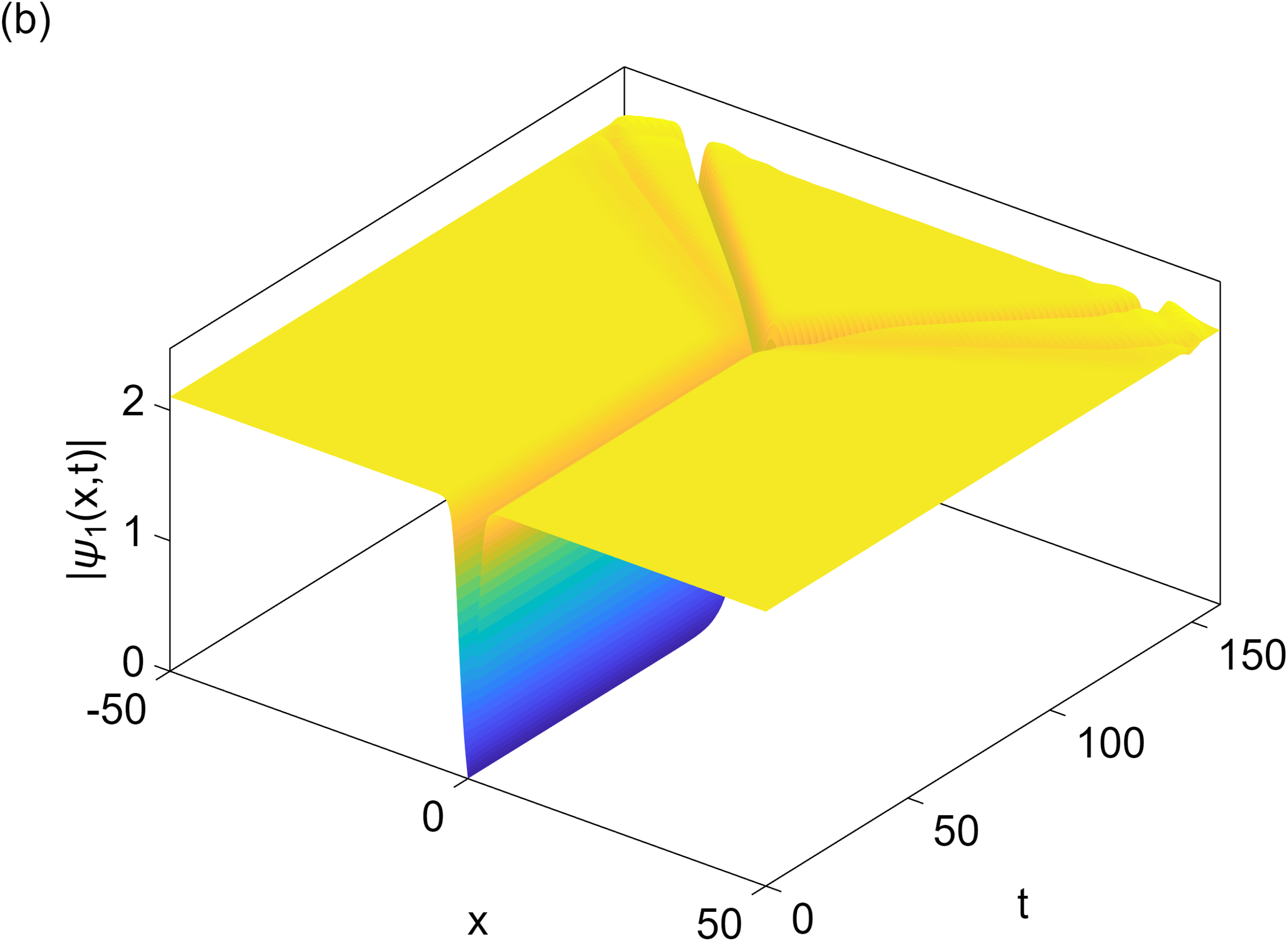}}
  \centerline{\includegraphics[width=4.45cm]{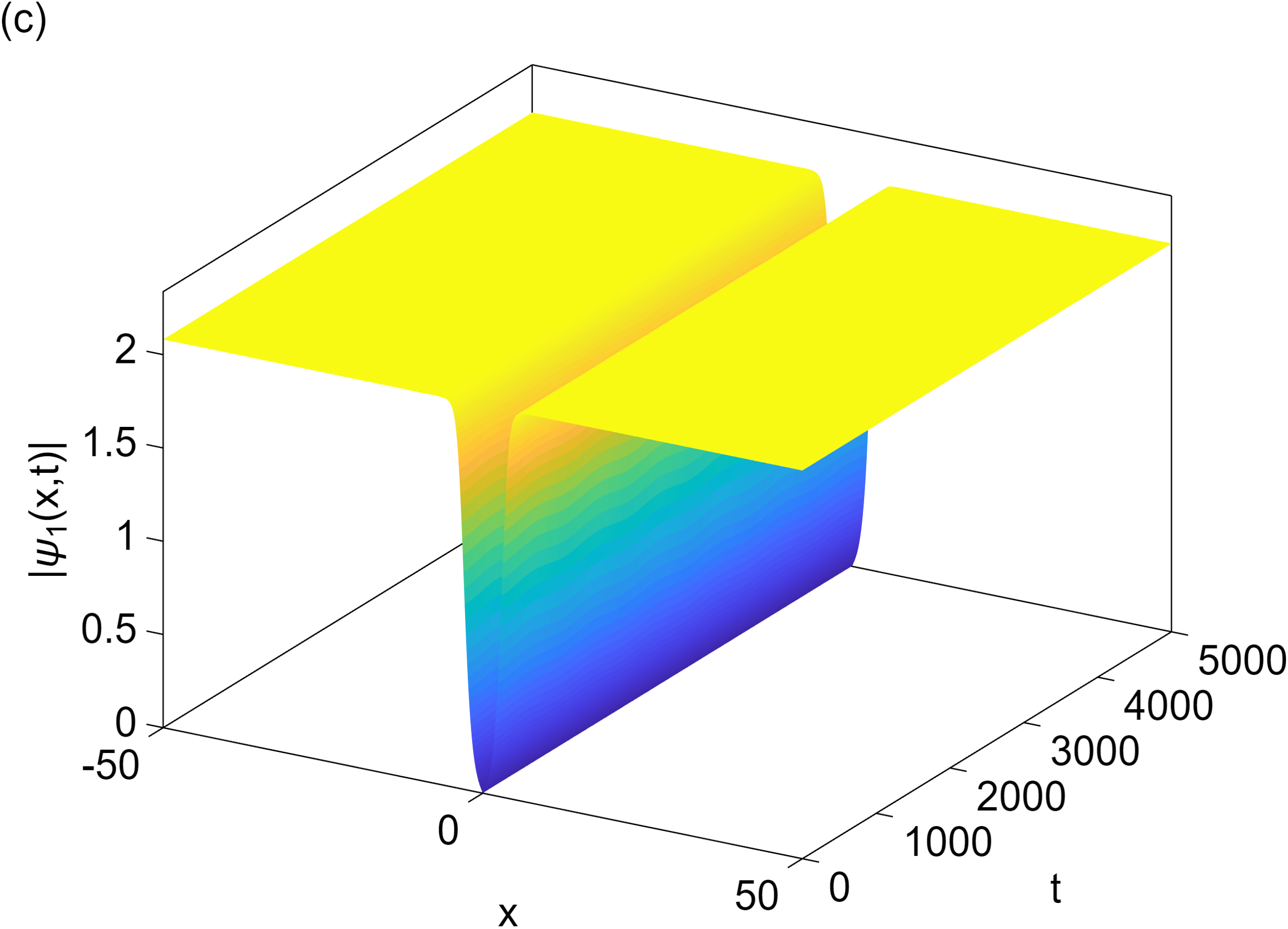}}
\caption{Time evolution of perturbed asymmetric hole solutions at fixed $\mu_2=-0.5$ for (a) $\mu_1=-0.4$, (b) $\mu_1=-0.9$, and (c) $\mu_1=-0.945$. Panels (a) and (b) show unstable dynamics, where the initial hole breaks into oppositely moving gray solitons, whereas panel (c) demonstrates stable evolution. The density dynamics are shown only for the first component, since the second component evolves in a qualitatively similar way. Different time scales are used in the three panels. The remaining parameters are $\sigma=10$, $\gamma=1$, $\delta=0.4$, and $\delta_1=\delta/2^{3/2}$. }
\label{fig-k2k}
\end{figure}

The modulational stability of the asymptotic background is necessary for a robust hole, but it is not sufficient to guarantee stability of the complete localized state. In a two-component system, perturbations may also excite an internal relative-density mode localized around the notch. Such a spinor mode can destabilize the hole even though the far-field upper background is modulationally stable. The resulting dynamics are shown in Fig.~\ref{fig-k2k}.

Figure~\ref{fig-k2k}(a) presents the evolution at $\mu_1=-0.4$ and fixed $\mu_2=-0.5$. The initially stationary hole is unstable and separates into two gray solitons moving in opposite directions. Panel~\ref{fig-k2k}(b), corresponding to $\mu_1=-0.9$, exhibits the same qualitative breakup. However, the longer characteristic time required for the instability to develop indicates that its growth becomes weaker as the cutoff is approached.

In Fig.~\ref{fig-k2k}(c), $\mu_1=-0.945$ lies close to $\mu_{1,\mathrm{co}}\simeq-0.94731$. In this regime, the broad kink-like hole preserves its form throughout the simulated interval and is dynamically stable. Only the first-component density is displayed in all three panels because the second component undergoes qualitatively similar dynamics. The different time scales used in panels (a)-(c) emphasize the progressive weakening of the instability near the coexistence limit.

The simulations therefore reveal a narrow stability region close to the kink coexistence point. Away from this limit, asymmetric holes are destabilized by an internal spinor mode and break into counterpropagating gray states. By contrast, sufficiently broad holes near $\mu_{1,\mathrm{co}}$ remain dynamically robust. This behaviour differs fundamentally from that of the scalar symmetric model, where the corresponding holes are stable throughout their existence domain, and demonstrates that the stability of a multicomponent localized excitation cannot be inferred solely from the modulational stability of its asymptotic background.

\begin{figure}[htbp]
\centerline{ \includegraphics[width=4.4cm]{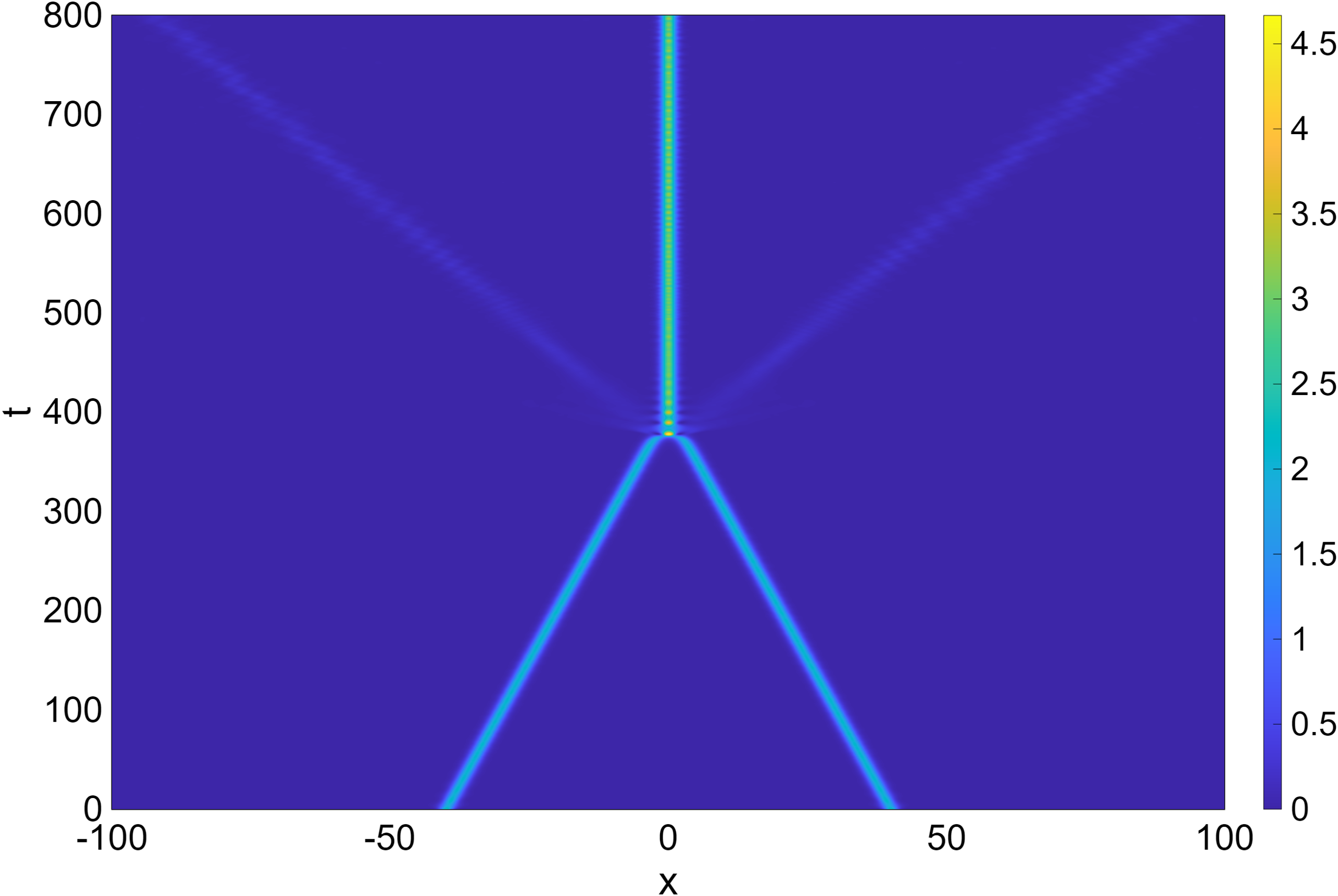} \hskip-0.1cm \includegraphics[width=4.4cm]{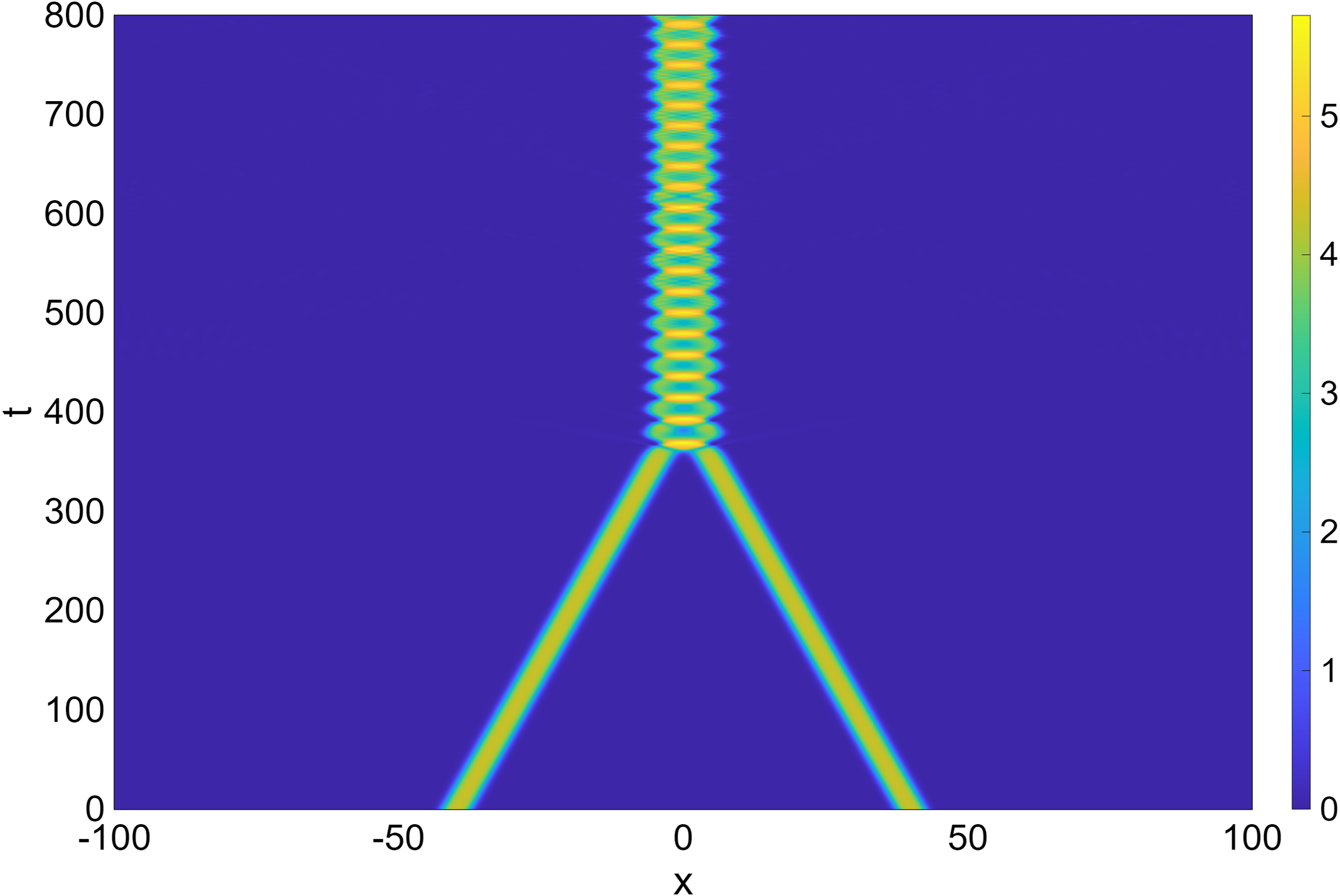}}
\centerline{ \includegraphics[width=4.4cm]{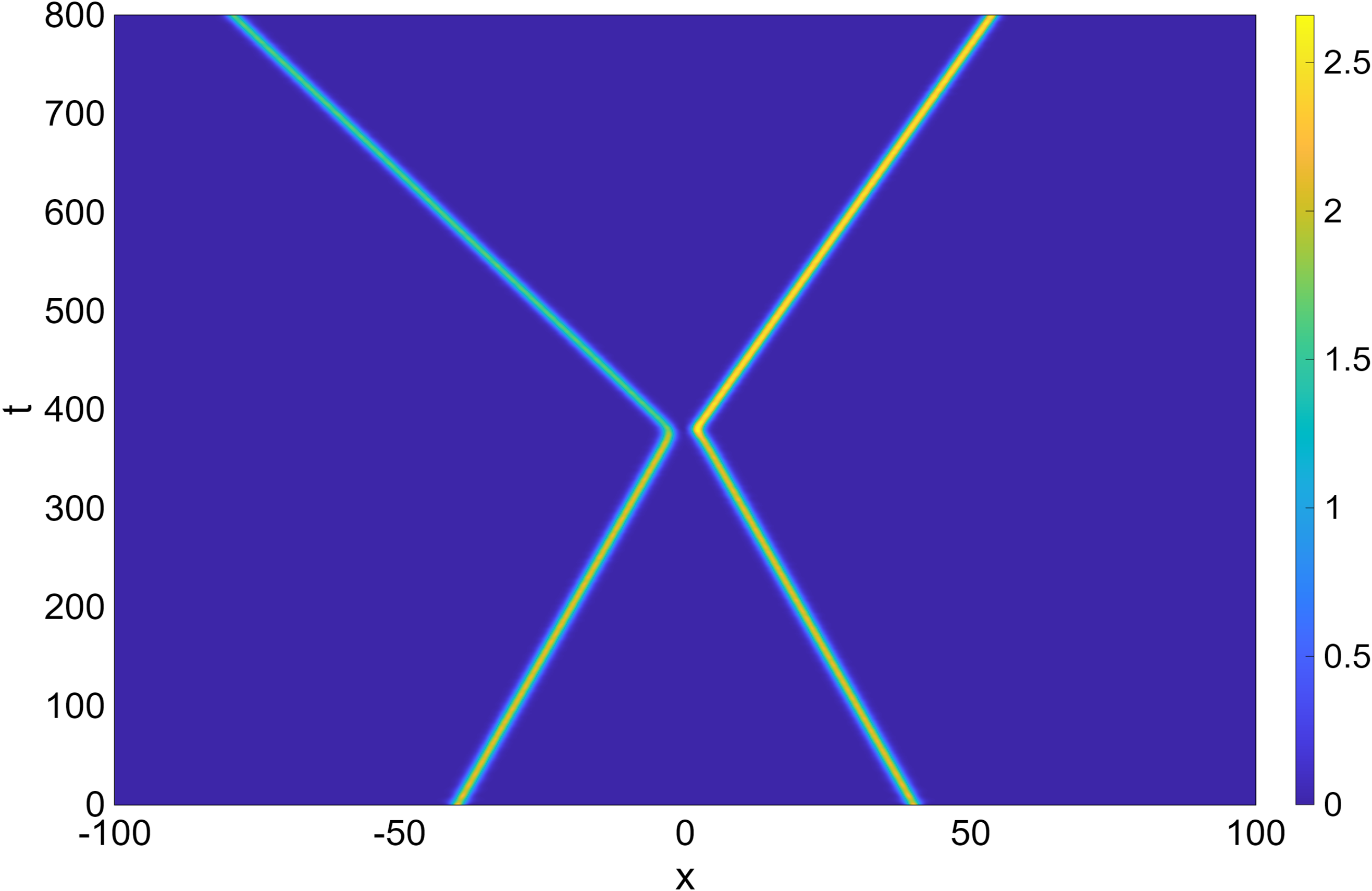} \hskip-0.1cm \includegraphics[width=4.4cm]{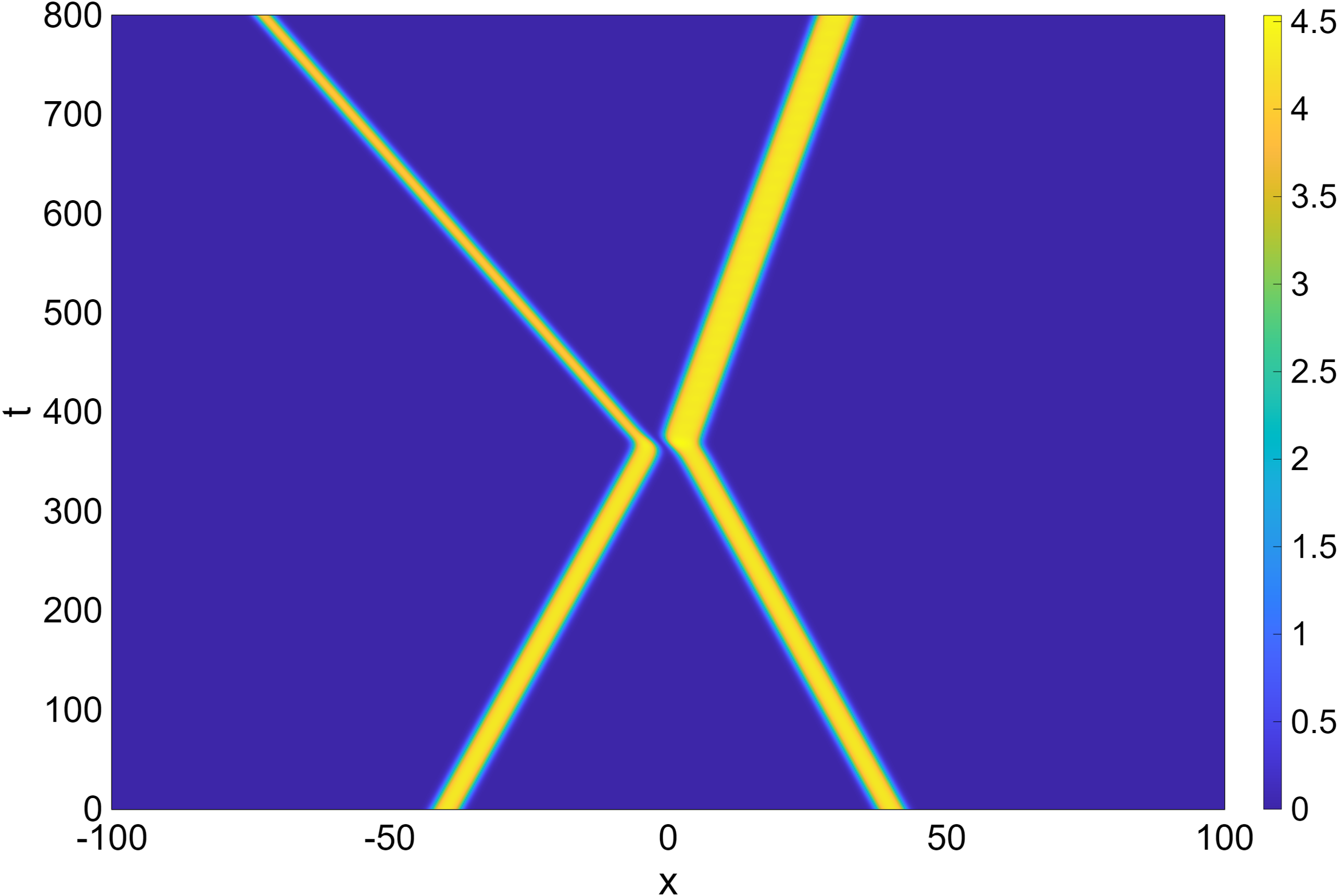}}
\centerline{ \includegraphics[width=4.4cm]{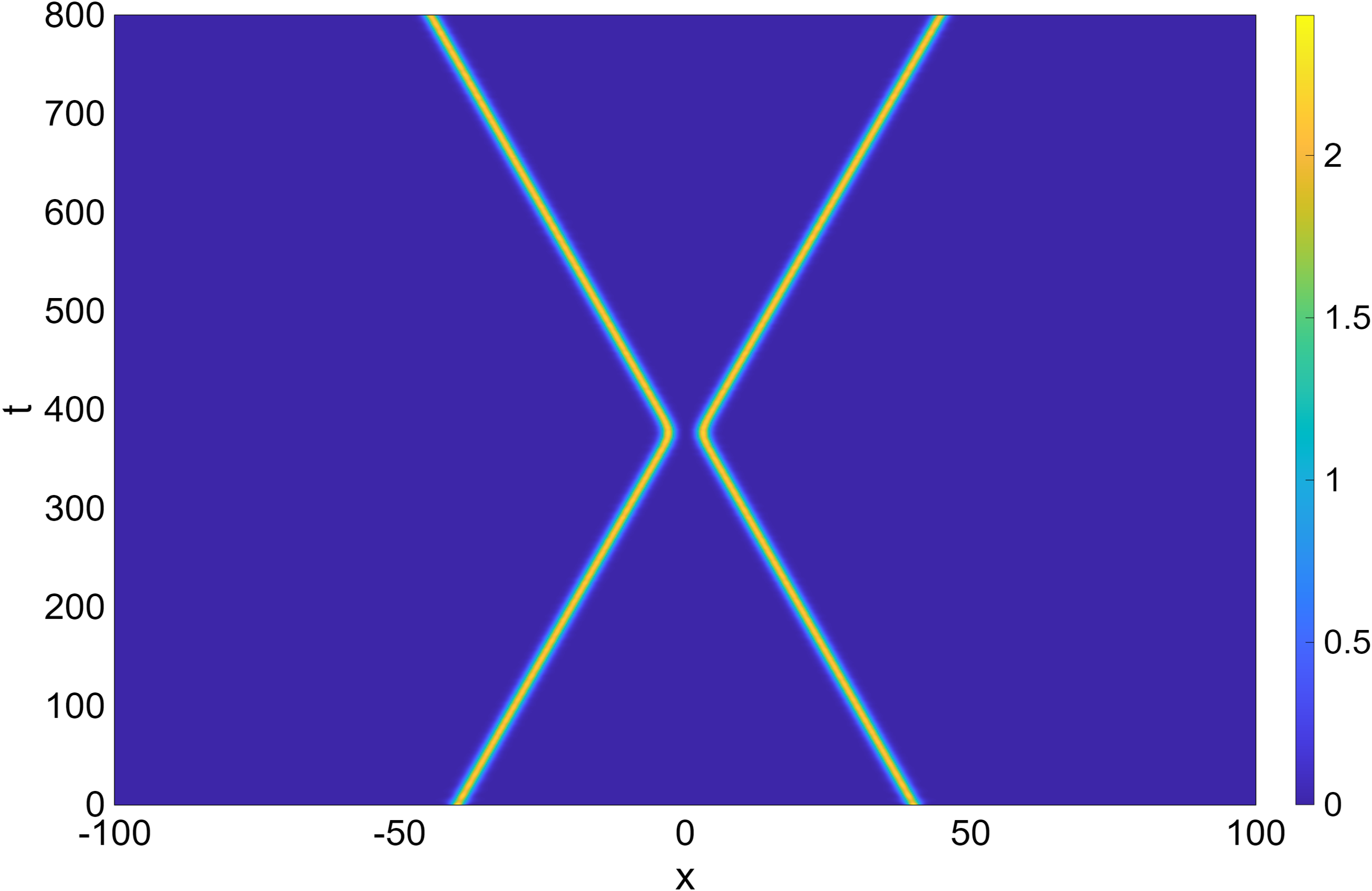} \hskip-0.1cm \includegraphics[width=4.4cm]{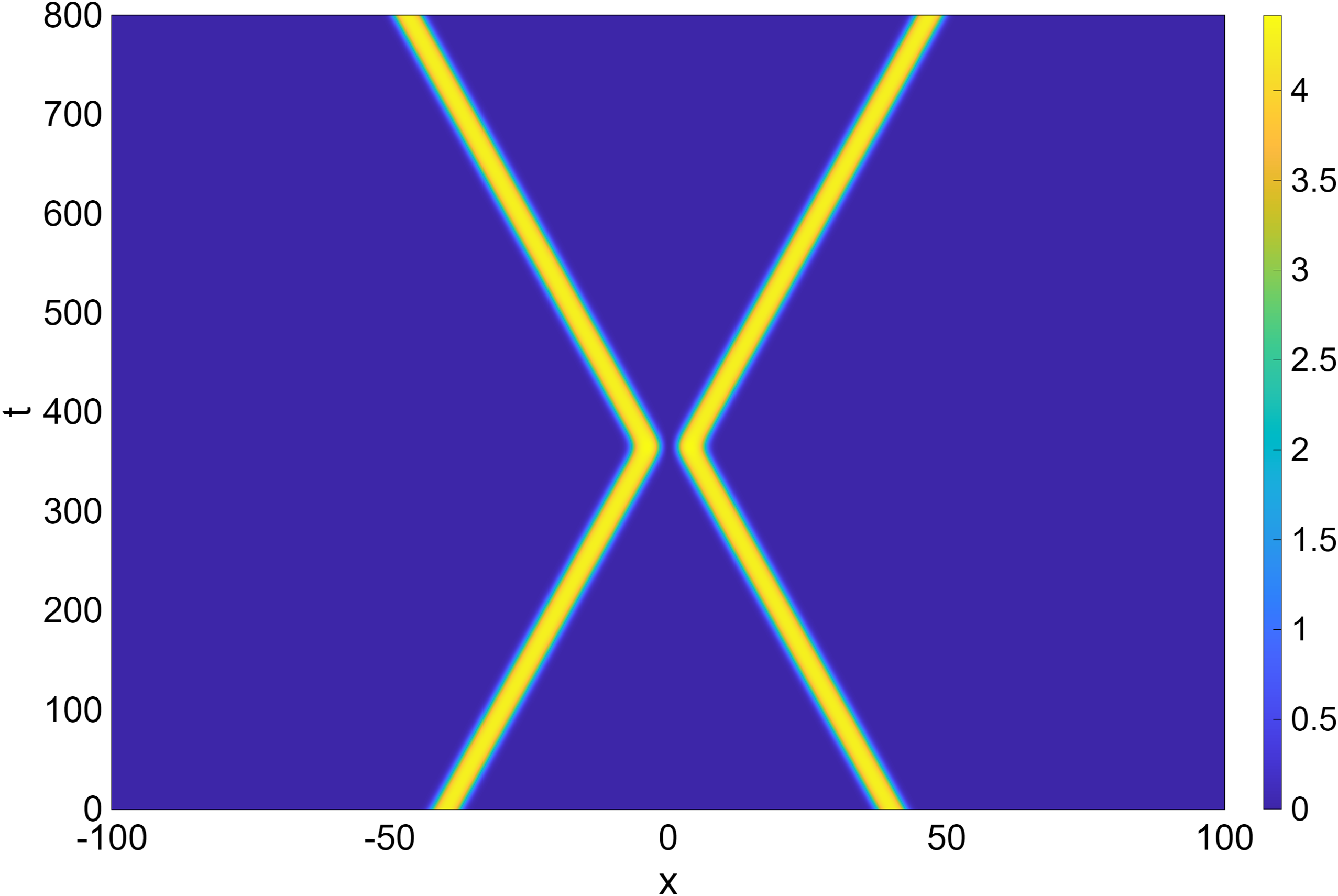}}
\caption{Collision dynamics of bright droplet-like localized states formed from kink-antikink superpositions. The left column displays compact bell-shaped states with internal separation $d_0=2$, while the right column presents broader flat-top states with $d_0=6$. In both cases, the two incident states are initially positioned at $x=\pm40$. The rows illustrate collisions for relative phases $\beta=0$, $\beta=\pi/2$, and $\beta=\pi$, respectively. In-phase states merge, intermediate-phase collisions result in asymmetric particle redistribution, and out-of-phase states show repulsion. The remaining parameters are $\delta g=1$ and $g_{\mathrm{LHY}}=0.4$.}
\label{fig:CollisionDynam}
\end{figure}

\subsection{Collision behavior of kink-antikink droplet-like states}
\label{sec:CollisionDynam}

We next investigate whether bright localized states constructed from kink-antikink superpositions reproduce the characteristic collision behavior of genuine self-bound quantum droplets. To do so, we solve the time-dependent Gross-Pitaevskii equation~(\ref{eq:gpe}) numerically using the split-step Fourier method with periodic boundary conditions. The computational domain is chosen sufficiently large so that the colliding states and the emitted low-amplitude waves do not interact with their periodic images over the considered time interval.

A single bright droplet-like state with internal kink-antikink separation $d_0$ is constructed using the multiplicative ansatz

\begin{equation}
\Phi_{\mathrm{KAK}}(x,d_0)
=
\frac{
q_{\mathrm{K}}\left(x+\dfrac{d_0}{2}\right)
q_{\mathrm{AK}}\left(x-\dfrac{d_0}{2}\right)
}{
A_{\mathrm{co}}
}.
\label{eq:single_KAK_collision}
\end{equation}

Here, $q_{\mathrm{K}}$ and $q_{\mathrm{AK}}$ denote the isolated kink and antikink profiles, respectively, and $A_{\mathrm{co}}$ represents the coexistence-background amplitude. Division by $A_{\mathrm{co}}$ ensures the correct plateau amplitude in the region between well-separated fronts. As a result, small $d_0$ produces a compact bell-shaped state through strong kink-antikink overlap, whereas larger $d_0$ yields a wider flat-top state with a larger particle number.

Two identical counterpropagating states are initially positioned at $x=-x_0$ and $x=+x_0$. The initial condition for the collision is given by

\begin{align}
\Psi(x,0)
={}&
\Phi_{\mathrm{KAK}}(x+x_0,d_0)
\exp\!\left[ik(x+x_0)\right]
\notag\\
&+
\Phi_{\mathrm{KAK}}(x-x_0,d_0)
\exp\!\left[-ik(x-x_0)+i\beta\right].
\label{eq:KAK_collision_IC}
\end{align}

Here, $k>0$ specifies the magnitude of the incident velocity, and $\beta$ denotes the initial relative phase. The left state carries momentum $+k$ and moves to the right, while the right state carries momentum $-k$ and moves to the left. The phase factors in Eq.~(\ref{eq:KAK_collision_IC}) therefore ensure a phase difference of exactly $\beta$ between the two droplet centers. Because the two incident states have equal norms and opposite momenta, the total initial momentum is zero. 

Figure~\ref{fig:CollisionDynam} presents a comparison of collisions for two different internal kink-antikink separations. The left column corresponds to $d_0=2$, where each incident state shows a compact bell-shaped profile and a relatively small particle number. The right column corresponds to $d_0=6$, giving broader flat-top states with larger norms. In both cases, the initial centers are located at $x=\pm40$, yielding an initial center-to-center separation of $80$. The three rows, from top to bottom, correspond to the relative phases $\beta=0$, $\beta=\pi/2$, and $\beta=\pi$.
In the in-phase case, $\beta=0$, the two states attract and merge after the collision, forming a heavier localized structure near the center of the domain. The merged state typically exhibits internal oscillations, as a portion of the incident kinetic energy is converted into breathing excitations. For the compact case $d_0=2$, weak radiation is emitted symmetrically from the collision region because the strongly overlapping kink-antikink ansatz represents only an approximate stationary state and is more compressible than the flat-top counterpart. By contrast, for $d_0=6$, the larger state approaches the incompressible flat-top regime, and the post-collision central structure is correspondingly broader. 

For the intermediate phase difference, $\beta=\pi/2$, the collision is asymmetric, even though the incident norms and velocities are initially equal. A phase dependent transfer of particles occurs during the overlap, resulting in two outgoing localized states with different norms. According to the phase convention in Eq.~(\ref{eq:KAK_collision_IC}), the right-moving outgoing state contains a larger fraction of the particles, while the left-moving state has a smaller norm. Their outgoing velocities also differ. Nevertheless, the total norm and total momentum of the wave field remain conserved, so the velocity difference compensates for the redistribution of particles between the two states.
In the out-of-phase case, $\beta=\pi$, destructive interference produces an effective repulsive interaction. The two localized states approach the collision region, decelerate, and then reverse direction, resulting in an evolution that resembles a mirror reflection. Because the initial configuration is symmetric and has zero total momentum, the reflected states propagate symmetrically away from the collision point. This repulsive response occurs in both the bell-shaped and flat-top configurations, although their distortion and radiation emission may differ due to differences in compressibility and particle number.

Overall, the phase dependent collision interactions of kink-antikink-based localized states are qualitatively identical to those observed in genuine self-bound quantum droplets: in-phase states merge, intermediate phase differences result in particle transfer and asymmetric scattering, and out-of-phase states show repulsion or reflection. The agreement between the two systems improves as $d_0$ increases, since the multiplicative kink-antikink ansatz converges to the exact flat-top quantum droplet profile in the large-separation limit. More pronounced radiation in the $d_0=2$ case mainly arises from the stronger overlap of the constituent fronts and the resulting deviation of the approximate initial profile from an exact stationary quantum droplet.

\section{Conclusions}
\label{sec:Conc}

In this work, we investigated the existence, structure, and stability of kinks, antikinks, holes, bubbles, and droplet-like states in an elongated Bose-Bose mixture with attractive residual mean-field interactions and repulsive Lee-Huang-Yang corrections. The homogeneous backgrounds have two constant-amplitude branches: the upper branch is modulationally stable, whereas the lower branch is unstable.

Using the energy, grand potential, and thermodynamic pressure, we showed that vacuum-finite-density coexistence selects a unique background on the stable upper branch. At this point, the finite-density pressure vanishes and a stationary kink or antikink connects the background to the vacuum. Real-time evolution confirms the robustness of these fronts against the applied perturbations over the simulated interval.
Symmetric holes form a continuous family of states embedded in the stable upper background. As coexistence is approached, their width and integrated density deficit increase without bound, connecting localized holes to a pair of increasingly separated fronts. The same fronts provide building blocks for droplet-like states: a normalized kink-antikink product approaches a broad flat-top stationary droplet as the separation increases. With the antikink to the left of the kink, their sum instead gives a bubble-like depletion without a phase jump, whereas their difference gives a dark-hole-like profile with a $\pi$ phase difference between the asymptotic backgrounds.

For bubble-like profiles, increasing the front separation deepens the depletion. The normalized depth distinguishes gray-like from dark-like bubbles but is not a stability criterion. A deep bubble persists under a small random perturbation over the simulated interval, establishing finite-time robustness rather than spectral stability. In the cubic-quadratic GPE, stationary bubbles remain spectrally unstable, although their instability growth rate decreases toward the deep, wide-bubble limit as the complementary atom number increases~\cite{Katsimiga2023}. Whether the present cubic-quartic model exhibits the same growth-rate dependence requires a Bogoliubov-de Gennes analysis of its stationary bubbles.
In the two-component system, perturbations separate into total-density and relative-density channels. Although the lower homogeneous branch is unstable through the total-density channel, stability of the far-field background alone does not guarantee stability of an asymmetric hole: localized relative-density modes can induce spinor-driven breakup. Finally, collisions of front-constructed droplet-like states show phase-dependent merging, asymmetric particle transfer, and effective repulsion.
Overall, these results connect homogeneous phases, nonlinear fronts, holes, bubbles, and self-bound droplets in a unified front-based picture.

\section{Acknowledgments}
This work has been supported by the State Budget of the Republic of Uzbekistan (Grant No. 2026 year award).

\end{document}